\documentclass[aps,prd,preprint,floats,superscriptaddress,nofootinbib]{revtex4}
\usepackage[utf8]{inputenc}
\usepackage{amsmath,amssymb}
\usepackage{graphicx}
\usepackage{color,slashed}
\pdfoutput=1

\begin{document}
\title{\bf Lepton flavor violation and scotogenic Majorana neutrino mass in a Stueckelberg $U(1)_X$ model}

\author{Chuan-Hung Chen}
\email[E-mail: ]{physchen@mail.ncku.edu.tw}
\affiliation{Department of Physics, National Cheng-Kung University, Tainan 70101, Taiwan}
\affiliation{Physics Division, National Center for Theoretical Sciences, Taipei 10617, Taiwan}

\author{Cheng-Wei Chiang}
\email[E-mail: ]{chengwei@phys.ntu.edu.tw}
\affiliation{Department of Physics and Center for Theoretical Physics, National Taiwan University, Taipei 10617, Taiwan}
\affiliation{Physics Division, National Center for Theoretical Sciences, Taipei 10617, Taiwan}

\author{Takaaki Nomura}
\email[E-mail: ]{nomura@scu.edu.cn}
\affiliation{College of Physics, Sichuan University, Chengdu 610065, China}

\author{Chun-Wei Su}
\email[E-mail: ]{r10222026@ntu.edu.tw}
\affiliation{Department of Physics and Center for Theoretical Physics, National Taiwan University, Taipei 10617, Taiwan}

\date{\today}

\begin{abstract}
We construct a scotogenic  Majorana neutrino mass model in a gauged $U(1)_X$ extension of the standard model, where the mass of the gauge boson and the unbroken gauge symmetry, which leads to a stable dark matter (DM), can be achieved through the Stueckelberg mechanism.  It is found that the simplest version of the extended model consists of the two inert-Higgs doublets and one vector-like singlet fermion. In addition to the Majorana neutrino mass, we study the lepton flavor violation (LFV) processes, such as $\ell_i \to \ell_j \gamma$, $\ell_i \to 3 \ell_j$, $\mu-e$ conversion rate in nucleus, and muonium-antimuonium oscillation. We  show that the sensitivities of $\mu\to 3e$ and $\mu-e$ conversion rate designed in Mu3e and COMET/Mu2e experiments make both decays the most severe constraints on the $\mu\to e$ LFV processes. It is found that $\tau\to \mu \gamma$ and $\tau\to 3\mu$ can reach the designed significance level of Belle II.  In addition to explaining the DM relic density, we also show that the DM-nucleon scattering cross section can satisfy the currently experimental limit of DM direct detection. 

\end{abstract}
\maketitle

%%%%%%%%%%%%%%%%%%%%%%%%%%%%%%%%%%%%%%%%%%%%%%%%%%%%%%%%%%%%
\section{Introduction}
%%%%%%%%%%%%%%%%%%%%%%%%%%%%%%%%%%%%%%%%%%%%%%%%%%%%%%%%%%%%

A scotogenic mechanism, which generate the neutrino mass radiatively, was proposed in Ref.~\cite{Ma:2006km} and  is referred to as the Ma model hereinafter.  The model not only resolves the origin of neutrino mass but also provides a way to explain the dark matter (DM) relic density, where the current value observed by Planck collaboration is $\Omega_{\rm DM} h^2= 0.120 \pm 0.001$~\cite{Planck:2018vyg}.

Since the active neutrinos and the charged leptons form $SU(2)_L$ doublets in the standard model (SM), the mechanism proposed to explain  the neutrino mass usually also induces interesting phenomena with lepton flavor violation (LFV), such as $\mu\to e\gamma$, $\mu\to 3e$, $\mu-e$ conversion in nucleus, and muonium-antimuonium oscillation, that are highly suppressed in the SM.

Due to their high sensitivities to new physics effects, many ongoing or planning experiments are designed to achieve an unprecedented precision to study the LFV processes.  For instance, the branching ratio (BR) for the $\mu \to e \gamma$ decay in MEG II experiment can reach the sensitivity of $6\times 10^{-14}$~\cite{MEGII:2018kmf};  $BR(\mu \to 3e) \lesssim 10^{-16}$ is expected in Mu3e experiment~\cite{Perez:2021gnr}; the $\mu-e$ conversion rate in COMET at J-PARC~\cite{COMET:2018auw} and Mu2e at FNAL~\cite{Diociaiuti:2020yvo}  can reach the level of $10^{-17}-10^{-18}$, whereas the  PRISM/PRIME experiment will push the conversion rate down below $10^{-18}$~\cite{Barlow:2011zza}; and  the  probability of  muonium-antimuonium conversion in the new generation experiment with the sensitivity of ${\cal O}(10^{-14})$ is proposed by the MACE collaboration at CSNC~\cite{MACE:CSNC}.

In order to explain the observed neutrino mass and DM relic density and to generate testable LFV processes in a scotogenic model, we study a gauged $U(1)_X$ extension of the SM, where only the newly introduced particles carry the $U(1)_X$ charges. The neutrino mass then arises from the quantum corrections where the particles carrying the $U(1)_X$ charges are running in the loops.  The DM candidate is stable if the  $U(1)_X$ gauge symmetry is unbroken.  However, the unbroken $U(1)_X$ would normally leads to a massless dark photon ($Z'$) and, thus, the DM relic density may not be correctly reproduced through the resonant $Z'$ production when we focus on the minimal extension of the SM. 
 %In addition, a massive $Z'$ can avoid the small $U(1)_X$ charge~\cite{Fabbrichesi:2020wbt}. 
To guarantee  the coexistence of a massive $Z'$ gauge boson and gauge invariance, we consider the Stueckelberg mechanism by introducing a Stueckelberg scalar field instead of the mechanism of spontaneous symmetry breaking.

Using the scotogenic model to radiatively generate the Dirac neutrino mass in the Stueckelberg mechanism  was studied in Ref.~\cite{Leite:2020wjl}.  In this work, using a different framework, we construct a model that can induce the Majorana neutrino mass.  To retain the $U(1)_X$ gauge anomaly-free,  we use a vector-like singlet Dirac fermion ($N_{L,R}$) instead of the right-handed singlet Majorana fermion in the Ma model.  Utilizing the unbroken $U(1)_X$ gauge symmetry to stabilize DM is done at the cost of introducing two inert-Higgs doublets.  Because the two inert-Higgs doublets ($\eta_{1,2}$) carry the lepton number and different $U(1)_X$ charges, lepton number violation (LNV) for generating the Majorana neutrino mass arises from the quartic coupling $(H^\dag \eta_1) ( H^\dag \eta_2)$ in the scalar potential.  In addition, since the  Yukawa couplings to the SM lepton can have different structures, such as $\overline L N_R \tilde\eta_1$ and $\overline{L^C} N_L \tilde\eta_2$, it is found that the neutrino data can be fitted with only one generation $N_{L,R}$.

When the neutrino oscillation data and the upper limits on LFV processes are taken into account, it is found that the $\mu-e$ conversion rate and the $\mu\to 3e$ decay will exclude most of the parameter space that lead to $BR(\mu\to e \gamma)\sim O(10^{-15})$, where the former is mediated by the photon-penguin diagrams and the letter is dominated by the box diagrams in the model, respectively.  In other words, both experiments will give the strongest constraints on new physics for the $\mu\to e$ LFV processes.  In addition, the resulting $BR(\tau\to \mu \gamma)$ and $BR(\tau\to 3\mu)$ can reach the significant levels of $O(10^{-9})$ and $O(10^{-10})$, respectively, as expected to probe by Belle II experiment~\cite{Belle-II:2018jsg}.

Two sources can contribute to the DM relic density in the model. One is through the Yukawa couplings and the other arises from the gauge coupling of the kinetic mixing.  The kinetic mixing between $Z'$ and the gauge field of $U(1)_Y$ can be induced via quantum corrections; thus, the $Z'$ gauge boson can couple to the SM particles.  The singlet fermion $N$, which is the DM candidate in the model, can then annihilate into the SM particles through the $Z'$-mediated s-channel process $NN\to Z' \to F \bar F$ and the $N$-mediated t-channel one $NN\to Z'(\to F \bar F) Z'(\to F' \bar F')$, where $F(F')$ denotes possible SM particles.  We will demonstrate that in addition to the Yukawa coupling scenario, the induced $Z'$ gauge coupling one can also explain the DM relic density $\Omega_{\rm DM} h^2$.  Moreover, the $Z'$ couplings to quarks will contribute to the DM-nucleon scattering.  It is found that the resulting DM-nucleon scattering cross section is under the current upper bound of direct search of DM from the XENON1T experiment~\cite{XENON:2018voc}.

The structure of this paper is organized as follows.  We introduce the model and derive the relevant scalar couplings, Yukawa couplings, and $Z'$ gauge couplings in Sec.~\ref{sec:model}.  In Sec.~\ref{sec:pheno}, we formulate the phenomena for neutrino mass and LFV processes.  Sec.~\ref{sec:NA} contains the detailed numerical analysis.  Finally, we summarize our findings of the study in Sec.~\ref{sec:summary}.

%%%%%%%%%%%%%%%%%%%%%%%%%%%%%%%%%%%%%%%%%%%%%%%%%%%%%%%%%%%%
\section{The Model} \label{sec:model}
%%%%%%%%%%%%%%%%%%%%%%%%%%%%%%%%%%%%%%%%%%%%%%%%%%%%%%%%%%%%

To obtain the Majorana mass for neutrinos through radiative corrections in a scotogenic $U(1)_X$ Stueckelberg gauge model, we find that a minimal extension of the SM is to include one singlet vector-like fermion $N$ and two new inert scalar doublets $\eta_{1,2}$, where their representations and charge assignments are given in Table~\ref{tab:rep}.  Note that here we use the convention that the electromagnetic charge $Q = T_L^3 + Y/2$.  To break the lepton number in the scalar sector,  $\eta_{1,2}$ need to carry one unit of lepton number while the singlet vector-like fermion has no lepton number.  As a result, the new particles are $R$-parity odd and the SM particles are $R$-parity even, where the $R$-parity quantum number is defined as $R_p=(-1)^{3B+ L + 2 S}$, with $B$, $L$, and $S$ bring the baryon number, lepton number, and spin of the particle, respectively.  Due to the odd $R$-parity property, $N$ and the neutral components in $\eta_{1,2}$ can be DM candidates.  Based upon the charge assignments, in the following we discuss the relevant Yukawa and gauge couplings for the phenomenological analysis.

%%%%%%%%%%%%%%%%%%%%%%%%%%%%%%%%%%%%%%%%%%%%%%%%%%%%%%%%%%%%
\begin{table}[thp]
 \caption{ Representations and charged assignments of new particles. }
\begin{center}
\begin{tabular}{cccccc} \hline \hline
  & ~~$SU(2)_L$~~ & ~~$U(1)_Y$~~  &  ~~$U(1)_X$~~ & ~~Lepton \#~~ & $R_p$  \\ \hline
$N$ & 1 & 0 & $Q_X$ & 0  & $-1$ \\ \hline
 $\eta_1$ & 2 & 1 & $Q_X$ & 1 & $-1$\\ \hline
 $\eta_2$ &  2 & 1 & $-Q_X$ & 1 & $-1$\\ \hline \hline  
\end{tabular}
\end{center}
\label{tab:rep}
\end{table}%
%%%%%%%%%%%%%%%%%%%%%%%%%%%%%%%%%%%%%%%%%%%%%%%%%%%%%%%%%%%%

%%%%%%%%%%%%%%%%%%%%%%%%%%%%%%%%%%%%%%%%%%%%%%%%%%%%%%%%%%%%
\subsection{Scalar masses and mixings, and Yukawa couplings}
%%%%%%%%%%%%%%%%%%%%%%%%%%%%%%%%%%%%%%%%%%%%%%%%%%%%%%%%%%%%

Apart from the Yukawa couplings, the most important effect to generate the Majorana mass from the scotogenic mechanism is the appearance of LNV term in the scalar potential, which dictates the scalar masses, couplings, and mixings.  To examine these effects in the model, we write the scalar potential for the SM Higgs $H$ and $\eta_{1,2}$ as:
\begin{align}
\begin{split}
 V =& V_{\rm SM} + V(H,\eta_1,\eta_2)~, 
 \\
 V_{\rm SM} =& - \mu^2_H H^\dag H + \lambda_H (H^\dag H)^2 ~,
 \\
 V(H, \eta_1,\eta_2 ) =& 
 \mu^2_1 \eta^\dag_1 \eta_1+ \mu^2_2 \eta^\dag_2 \eta_2 + \lambda_1 (\eta^\dag_1 \eta_1)^2 + \lambda_2 (\eta^\dag_2 \eta_2)^2+ \lambda_3 (\eta^\dag_1 \eta_1) (\eta^\dag_2 \eta_2) 
 \\
 &+ \lambda_4 (\eta^\dag_1 \eta_2) (\eta^\dag_2 \eta_1)  + \left( \lambda_5 (H^\dag \eta_1) (H^\dag \eta_2) + \mbox{H.c.}\right) 
 + \lambda_6 (H^\dag \eta_1) (\eta^\dag_1 H) 
 \\
 &+ \lambda_7  (H^\dag \eta_2) (\eta^\dag_2 H)  +  \lambda_8 (H^\dag H) (\eta^\dag_1 \eta_1) +  \lambda_9 (H^\dag H) (\eta^\dag_2 \eta_2) ~.
\end{split}
\end{align}
It can be seen that  the only non-self-Hermitian term comes from the $\lambda_5$ term, which violates the lepton number by two units and  plays an important role on the radiative generation of the Majorana neutrino mass. The tiny neutrino mass can be achieved when $\lambda_5 \ll 1$, which is the same as that shown in the Ma model~\cite{Ma:2006km}.  For spontaneously breaking the  electroweak gauge symmetry, we take $\mu^2_H, \lambda_H >0$ as in the SM.  The masses of $\eta_{1,2}$ can be irrelevant to the electroweak symmetry breaking, and we thus  require $\mu^2_{1,2}(\lambda_{1,2}) > 0$.  To preserve the $U(1)_X$ and $R_p$ symmetries,  the vacuum expectation values (VEVs) of $\eta_{1,2}$ have to vanish. We therefore parametrize the components of the three doublet scalars as:
\begin{equation}
  H= 
\left(
\begin{array}{c}
  G^+     \\
  \frac{1}{\sqrt{2}} ( v + h + i G^0)      \\   
\end{array}
\right)~, ~~  \eta_j= 
\left(
\begin{array}{c}
  \eta^+_j    \\
  \frac{1}{\sqrt{2}} (s_j + i a_j)    \\   
\end{array}
\right)~, \label{eq:H_eta_rep}
\end{equation}
where $G^{\pm,0}$ are the Goldstone bosons, $v$ is the VEV of $H$, and  $h$ is the SM Higgs boson.

Since $\eta_1$ and $\eta_2$ carry different $U(1)_X$ charges, the charged scalars $\eta^\pm_1$ and $\eta^\pm_2$ do not mix and their masses are respectively obtained as:
\begin{align}
\begin{split}
  m^2_{\eta^\pm_1} & = \mu^2_1 + \frac{\lambda_8 v^2}{2}~,
  \\
  m^2_{\eta^\pm_2} & = \mu^2_2 + \frac{\lambda_9 v^2}{2}~.
\end{split}
\end{align}
Unlike the charged scalars, the neutral components of $\eta_{1,2}$ can mix via the $\lambda_5$ term.  According to the scalar potential with Eq.~(\ref{eq:H_eta_rep}), the mass-square matrices for $(s_1, s_2)$ and $(a_{1}, a_2)$ are respectively:
\begin{equation}
 m^2_{S} = \left(
\begin{array}{cc}
  m^2_{s_1} &    m^2_{12}  \\
 m^2_{12}   &   m^2_{s_2} \\   
\end{array}
\right)~, ~~ m^2_{A} =\left(
\begin{array}{cc}
  m^2_{s_1} &    - m^2_{12}  \\
 - m^2_{12}   &   m^2_{s_2} \\   
\end{array}
\right)~,
\end{equation}
with
\begin{align}
 m^2_{s_1} & = \mu^2_{1} + \frac{v^2}{2} (\lambda_6 + \lambda_8) ~, \nonumber \\
 m^2_{s_2} & = \mu^2_2 + \frac{v^2}{2} \left( \lambda_7 + \lambda_9 \right)~, \nonumber \\ 
 m^2_{12} & = \frac{v^2}{2} \lambda_5 ~.
\end{align}
Each of the two $2\times 2$ mass-square matrices can be diagonalized by the corresponding orthogonal matrix $O(\theta_\xi)$ ($\xi= S, A$) through $O(\theta_\xi) m^2_{\xi} O^T(\theta_\xi)$, where the $O(\theta_\xi)$ matrix can be parametrized as:
  \begin{equation}
  O(\theta_\xi) = \left(
\begin{array}{cc}
 \cos\theta_\xi &   \sin\theta_\xi  \\
 -\sin\theta_\xi  &   \cos\theta_\xi \\   
\end{array}
\right)~. \label{eq:omatrix}
  \end{equation}
Since the matrix elements in $m^2_A$ are the same as those in $m^2_S$ except the sign change in the off-diagonal elements, we therefore take $\theta_S=- \theta_A \equiv \theta$.  The eigenvalues for $m^2_S$ are found to be:
\begin{equation}
m^2_{S_{1,2}} = \frac{1}{2} \left[ m^2_{s_1} + m^2_{s_2} \pm \sqrt{(m^2_{s_2} - m^2_{s_1} )^2 - 4 (m^4_{12})} \right]~.
\end{equation}
For the physical pseudoscalars $A_{1,2}$, we have $m^2_{A_{1(2)}}= m^2_{S_{1(2)}}$.  The mixing angle $\theta$ is given by:
\begin{equation}
\sin(2\theta) = - \frac{\lambda_5 v^2}{m^2_{S_2} - m^2_{S_1}}~.
\end{equation}
Because $S_i$ and $A_i$ are degenerate in mass, if one of them is the DM candidate, the large gauge interaction $A_i-S_i-Z$ will render too large a DM-nucleon scattering cross section.  Hence, the possibility of using a scalar as the DM candidate in this model is excluded by the direct detection experiments.  Instead, the singlet vector-like fermion $N$ becomes a promising DM candidate in the model.  Since $\lambda_5$ term violates the lepton number and eventually leads to the Majorana mass, its value has to be sufficiently small, $\lambda_5 \ll 1$~\cite{Ma:2006km}, as alluded to earlier.  As a result, the off-diagonal mass matrix element $|m_{12}^2|$ is suppressed and the mixing angle $\theta\ll 1$.  In order to make the Yukawa couplings sufficiently large so that the LFV processes can be possibly detectable in the ongoing and planning experiments, we follow the approach in~\cite{Toma:2013zsa} and take $\lambda_5=10^{-9}$.

According to the $U(1)_X$ charges listed in Table~\ref{tab:rep}, the Yukawa couplings of $\eta_{1,2}$ and $N$ to the SM leptons are given by:
 \begin{align}
 -{\cal L}_Y &= 
 \overline L \, {\bf y}_1 \, \tilde\eta_1 N_{ R} + \overline L\, {\bf y}_2\, \tilde\eta_2 N^C_{ L } + m_{N} \overline N_{L} N_{R} 
 + \mbox{H.c.}~, \label{eq:yukawa}
 \end{align}
where the flavor indices are suppressed, $L^T=(\nu_\ell ,\ell )$ denotes the left-handed lepton doublet in the SM, ${\tilde\eta_j} = i \tau_2 \eta^*_j$ with $\tau_2$ being a Pauli matrix, $N^C= C \gamma_0 N^*$ is the charge conjugation of $N$, and $m_N$ is the mass of $N$.  Although ${\bf y}_{1,2}$ can be generally complex, we can rotate away the phase of one of them by redefining the phases of the complex lepton doublet $L$.  In our following analysis, we take the convention that ${\bf y}_1$ is real and ${\bf y}_2$ is complex, as parametrized by $y_{2 k} = |y_{2 k}| e^{i\phi_k}$ ($k = 1,2,3$).  Using Eqs.~(\ref{eq:H_eta_rep}) and $(\ref{eq:omatrix})$,  the Yukawa couplings in Eq.~(\ref{eq:yukawa})  can be decomposed as:
\begin{align}
\begin{split}
 -{\cal L}_Y =& 
 - \overline\ell_L  {\bf y}_1 N_R \eta^-_1 -   \overline\ell_L  {\bf y}_2 N^C_L \eta^-_2  + m_N \overline N_L N_R 
 \\
 & + \frac{1}{\sqrt{2}}\overline \nu_{\ell L}  {\bf y}_1 N_R  \left[ c_\theta (S_1 - i A_1) -s_\theta (S_2 + i A_2) \right] 
 \\
 &+ \frac{1}{\sqrt{2}}\overline \nu_{\ell L}  {\bf y}_2 N^C_L \left[ s_\theta (S_1 + i A_1) + c_\theta (S_2 - i A_2)\right] 
 + \mbox{H.c.}~, \label{eq:yuka}
\end{split}
\end{align}
where we denote $c_\theta~ (s_\theta) \equiv \cos\theta~ (\sin\theta)$. 
We note that  if we take the charged lepton $\ell_L$ as the mass eigenstate, in general, the $\eta^\pm_{1,2}$ Yukawa couplings are modified as $V^\ell_L {\bf y}_i$, where $V^\ell_L$ is a unitary matrix introduced for diagonalizing the charged lepton mass matrix.  Therefore, the Yukawa couplings of $\eta^\pm_i$ are generally different from those of $S_i$ and $A_i$.  Since we know nothing about the information of $V^\ell_L$ in the SM, we adopt $V^\ell_L=1$ to reduce the number of free parameters in the study.  As a result, the effects contributing to the neutrino mass through the Yukawa couplings of $S_i$ and $A_i$ now have a strong correlation to the LFV processes that are mediated by  $\eta^\pm_i$.

%%%%%%%%%%%%%%%%%%%%%%%%%%%%%%%%%%%%%%%%%%%%%%%%%%%%%%%%%%%%
\subsection{$U(1)_X$ Stueckelberg gauge couplings}
%%%%%%%%%%%%%%%%%%%%%%%%%%%%%%%%%%%%%%%%%%%%%%%%%%%%%%%%%%%%

The Lagrangian invariant under the $SU(2)_L\times U(1)_Y \times U(1)_X$ gauge symmetry with the Stueckelberg gauge field included is given by:
\begin{align}
\begin{split}
 {\cal L}_{\rm kin} =&
 - \frac{1}{4} \vec{ W}_{\mu\nu} \cdot \vec{W}^{\mu\nu} 
 - \frac{1}{4} \hat{B}_{\mu \nu} \hat{B}^{\mu\nu}
 - \frac{1}{4} \hat{Z'}_{\mu \nu} \hat{Z'}^{\mu \nu} 
 + \frac{m^2_{Z'}}{2} \left( \hat{Z'}^{\mu} - \frac{1}{m_{Z'}} \partial^\mu B \right)^2 
 \\
 & - \frac{1}{2 \beta } \left( \partial_\mu \hat{Z'}^{\mu} + m_{Z'} \beta B \right)^2
 + \overline N i \slashed {D} N + (D_\mu \eta_1)^\dag (D^\mu \eta_1)+ (D_\mu \eta_2)^\dag (D^\mu \eta_2)~,
 \label{eq:St}
\end{split}
\end{align}
where $W^i_{\mu \nu}$, $\hat{B}_{\mu\nu}$, and $\hat{Z'}_{\mu\nu}$ are the gauge field stress tensors associated with the  $SU(2)_L$, $U(1)_Y$, and $U(1)_X$ gauge groups, respectively, $B$ field is the Stueckelberg scalar field,  $\beta$ is the gauge fixing parameter, and $m_{Z'}$ is the mass of $U(1)_X$ Stueckelberg gauge field.  We note that the kinetic mixing term $\hat{B}_{\mu \nu} \hat{Z'}^{\mu \nu}$ can be rotated away by redefining the gauge fields $\hat{B}_\mu$ and $\hat{Z'}_\mu$.  However, we will show later that the mixings in $\gamma$-$Z'$ and $Z$-$Z'$ can be induced through radiative corrections.

The covariant derivatives on $N$ and $\eta_i$ are expressed as:
\begin{align}
\begin{split}
  D_\mu N &= (\partial_\mu + i g_X q_N \hat{Z'}_\mu) N ~, 
  \\
  D_\mu \eta_i & = \left(\partial_\mu +  i \frac{g}{2} \vec{\tau} \cdot \vec W_\mu + i \frac{g'}{2} \hat{B}_\mu+ i g_X q_{\eta_i} \hat{Z'}_\mu \right) \eta_i~,
\end{split}
\end{align}
 where  $q_{N( \eta_i)}$ denotes the $U(1)_X$ charge of $N(\eta_i)$, and $g$ and $g'$ are the gauge couplings of $SU(2)_L$ and  $U(1)_Y$, respectively. The $NNZ'$ interaction can be immediately obtained as:
  \begin{equation}
  {\cal L}_{NNZ'} = - g_X q_N  \overline N \gamma^\mu N \hat{Z'}_\mu~. 
  \end{equation} 
Although the quartic gauge couplings of $\eta_i$ to gauge bosons $W^\pm$, $Z$, $\gamma$, and $\hat{Z'}$ can be derived from the kinetic terms of $\eta_i$, these couplings are irrelevant to our study here and are not presented explicitly.  The various trilinear gauge couplings of $\eta_i$ from $(D_\mu \eta_i)^\dag (D^\mu \eta_i)$ can be summarized as:
\begin{align}
\begin{split}
 {\cal L}_{\rm kin} \supset&~ 
 i \frac{g}{\sqrt{2}}\left\{ \left[\partial_\mu \eta^-_i (S_i + iA_i)  - \eta^-_i ( \partial_\mu S_i + i \partial_\mu A_i)  \right] W^{+\mu} + \mbox{H.c.} \right\}
 \\
 & + \left(\frac{g}{2 c_W} Z^\mu - g_X q_X \hat{Z'}^{\mu}\right) \left[ (\partial_\mu S_i ) A_i - S_i \partial_\mu A_i \right] 
 \\
 & + i \left( e A^\mu + \frac{1-2 s^2_W}{2 c_W} g Z^\mu + g_X q_X \hat{Z'}^{\mu} \right) 
 \left[ (\partial_\mu \eta^{-}_i) \eta^+_i - \eta^-_i \partial_\mu \eta^+_i \right]~.
 \label{eq:gauge_kin}
\end{split} 
\end{align}   
The $W^\pm$, $Z$, and $\gamma$ gauge bosons and the Weinberg angle $\theta_W$ are defined through the relations:
\begin{align}
\begin{split}
  W^\pm &=  \frac{1}{\sqrt{2}} (W^1_\mu \mp W^2_\mu) ~,
  \\
 A_\mu &= c_W \hat{B}_\mu + s_W W^3_\mu~,
  \\
 Z_\mu &= -s_W \hat{B}_\mu + c_W W^3_\mu~,
\end{split}
\end{align}
where $c_W~(s_W) \equiv \cos\theta_W~(\sin\theta_W)$ and $e=g s_W = g' c_W$ are used.

In the study of LFV processes, we need to know the photon and $Z$-boson gauge couplings to the SM particles.  Therefore, the relevant photon and $Z$ gauge couplings are given by:
 \begin{equation}
 {\cal L}_{Z} = 
 - e Q_f \overline f \gamma_\mu f  A^\mu 
 - \frac{g}{2c_W} \overline f \left( C^f_R P_R + C^f_L P_L\right) \gamma_\mu f  Z^\mu~, 
 \label{eq:Z}
 \end{equation}
 with 
 \begin{equation}
 C^f_R = -2 Q_f s^2_{W}~, ~~ C^f_L = 2 T^3_f - 2Q_f s^2_{W}~,
 \end{equation}
where $Q_f$ is the electric charge of the fermion $f$, and $T^3_f$ is its third component of the weak isospin.

%%%%%%%%%%%%%%%%%%%%%%%%%%%%%%%%%%%%%%%%%%%%%%%%%%%%%%%%%%%%
\section{Neutrino mass, lepton-flavor violation, kinetic mixing of gauge boson} \label{sec:pheno}
%%%%%%%%%%%%%%%%%%%%%%%%%%%%%%%%%%%%%%%%%%%%%%%%%%%%%%%%%%%%

Based on the introduced interactions, in this section we investigate the new physics effects on the rare lepton flavor related processes, such as neutrino mass, $\ell_i \to \ell_j \gamma$, $\ell_i \to 3 \ell_j$, $\mu-e$ conversion in nuclei, and muonium-antimuonium oscillation.  Note that throughout this section, we assume that the new physics effects dominate and ignore the SM contributions. 

%%%%%%%%%%%%%%%%%%%%%%%%%%%%%%%%%%%%%%%%%%%%%%%%%%%%%%%%%%%%
\subsection{Scotogenic neutrino mass matrix}
%%%%%%%%%%%%%%%%%%%%%%%%%%%%%%%%%%%%%%%%%%%%%%%%%%%%%%%%%%%%

%%%%%%%%%%%%%%%%%%%%%%%%%%%%%%%%%%%%%%%%%%%%%%%%%%%%%%%%%%%%
\begin{figure}[phtb]
\begin{center}
\includegraphics[scale=1.2]{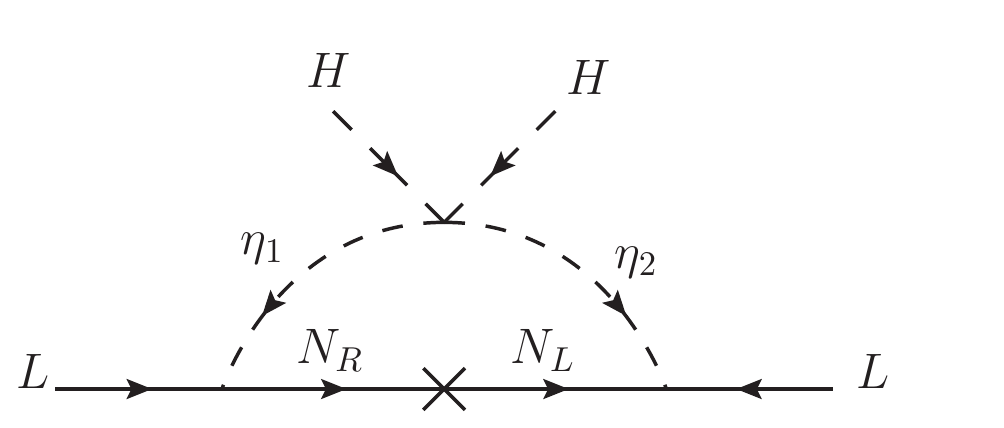}
\caption{Sketched Feynman diagram mediated by $Z_2$-odd particles for the  radiative  neutrino mass. }
\label{fig:loop_Feyn}
\end{center}
\end{figure}
%%%%%%%%%%%%%%%%%%%%%%%%%%%%%%%%%%%%%%%%%%%%%%%%%%%%%%%%%%%%

To obtain the Majorana neutrino mass in the model, we need both Yukawa couplings ${\bf y}_1$ and ${\bf y}_2$, which can ensure that the structure of  Majorana mass term  $L^T C L$ can be achieved. In addition, since  $\eta_{1,2}$ have no VEVs and the lepton number is preserved in the ground state, we are left with the choice of using the lepton number-violating interaction $\lambda_5 (H^\dag \eta_1) (H^\dag \eta_2)$ to generate the Weinberg's dimension-5 operator $L H H L$~\cite{Weinberg:1979sa}. For the purpose of clarity, we show a representative one-loop Feynman diagram in Fig.~\ref{fig:loop_Feyn}.  According to the Yukawa couplings in Eq.~(\ref{eq:yuka}), the neutrino mass matrix elements can be obtained as:
\begin{equation}
m^\nu_{ij}  = \frac{\sin(2\theta)}{16\pi^2}  \overline y_{ij} m_{N} \left[ \frac{m^2_{S_1}}{m^2_{S_1} -m^2_{N}} \ln\left( \frac{m^2_{S_1}}{m^2_{N} }\right) - \frac{m^2_{S_2}}{m^2_{S_2} -m^2_{N}} \ln\left( \frac{m^2_{S_2}}{m^2_{N} }\right) \right]~,  \label{eq:nu_mass}
\end{equation}
where we have included $A_i$ contributions,  $m_{A_i}=m_{S_i}$ is used, and the symmetric Yukawa couplings $\overline y_{ij}$  in flavor indices are defined as:
\begin{equation}
\overline y_{ij} =  y^{*}_{1i} y^{*}_{2j} + y^{*}_{2i} y^{*}_{1j}~.
\end{equation}
In the Ma model, the involved Yukawa couplings for $m^\nu_{ij}$ appear in the combination of $y_i y_j$; thus, the mass matrix elements have strong correlations.  It needs at least two singlet right-handed fermions to explain the neutrino data.  In our model, due to the introduction of one more inert Higgs doublet $\eta_2$, the induced neutrino mass matrix elements appear in the combination of $y_{1i} y_{2j} + y_{2i} y_{1j}$ and have less correlations among the matrix elements.  It is for this reason that the neutrino data can be accommodated using only one singlet fermion.

The symmetric mass matrix $m^\nu$ can be diagonalized through $m^{\nu}_{\rm dia}= U^\dag m^\nu  U^*$, where  $m^\nu_{\rm dia} ={\rm dia}(m_1, m_2, m_3)$ is the diagonal mass matrix. The unitary matrix $U$ can be parametrized as:
 \begin{equation}
  U= 
\left(
\begin{array}{ccc}
c_{12} c_{13}  &  s_{12} c_{13} &   s_{13} e^{-i \delta_{\rm CP}} \\
  -s_{12} c_{23} - c_{12} s_{13} s_{23} e^{i \delta_{\rm CP}}  
  &  c_{12} c_{23} - s_{12} s_{13} s_{23} e^{i \delta_{\rm CP}}  
  &  c_{13} s_{23}   \\
  s_{12} s_{23} - c_{12} s_{13} c_{23} e^{i\delta_{\rm CP}} 
  &  -c_{12} s_{23} - s_{12} s_{13} c_{23} e^{i \delta_{\rm CP}} 
  &   c_{13} c_{23}  
\end{array}
\right) \left(
\begin{array}{ccc}
 e^{i \alpha_1} &   &   \\
  &   e^{i \alpha_2} &   \\
  &   &   1
\end{array}
\right)~,  \nonumber
 \end{equation}
where $c_{ij}~(s_{ij}) \equiv \cos\theta_{ij}~(\sin\theta_{ij})$, $\delta_{\rm CP}$ is the Dirac CP-violating phase, and $\alpha_{1,2}$ are the Majorana phases.  Consequently, the theoretical mass matrix obtained in Eq.~(\ref{eq:nu_mass}) can be determined by the observables from the neutrino oscillations, and their relations are given by:
\begin{align}
\begin{split}
(m^\nu)_{11} =&~
m_{1} c^2_{12} c^2_{13} e^{i 2\alpha_1} + m_{2} s^2_{12} c^2_{13} e^{i 2\alpha_2} + m_3 s^2_{13} e^{-2 i \delta_{\rm CP}}  ~,
\\
(m^\nu)_{12} =&~
m_1 e^{2 i \alpha_1} c_{12} c_{13} \left( -s_{12} c_{23} - e^{i \delta_{\rm CP}} c_{12} s_{13} s_{23} \right) 
\\
& + m_2 e^{2 i \alpha_2} s_{12} c_{13} \left( c_{12} c_{23} - e^{i \delta_{\rm CP}} s_{12} s_{13} s_{23} \right) 
 + m_3 e^{-i \delta_{\rm CP}} s_{13} c_{13} s_{23}~,
 \\
 (m^\nu)_{13} =&~ 
 m_1 e^{2 i \alpha_1} c_{12} c_{13} \left( s_{12} s_{23} - e^{i \delta_{\rm CP}} c_{12} s_{13} s_{23} \right) 
 \\
 & + m_2 e^{2 i \alpha_2} s_{12} c_{13} \left(- c_{12} s_{23} - e^{i \delta_{\rm CP}} s_{12} s_{13} c_{23} \right) 
 + m_3 e^{-i \delta_{\rm CP}} s_{13} c_{13} c_{23}~, 
 \\
  (m^\nu)_{22} =&~ 
  m_1 e^{2 i \alpha_1}  \left( -s_{12} c_{23} - e^{i \delta_{\rm CP}} c_{12} s_{13} s_{23} \right)^2 
  \\
&  + m_2 e^{2 i \alpha_2} \left( c_{12} c_{23} - e^{i \delta_{\rm CP}} s_{12} s_{13} s_{23} \right)^2 
 + m_3   c^2_{13} s^2_{23}~, 
 \\
 (m^\nu)_{23} =&~
 m_1 e^{2 i \alpha_1}  \left( s_{12} s_{23} - e^{i \delta_{\rm CP}} c_{12} s_{13} c_{23} \right)  \left( -s_{12} c_{23} - e^{i \delta_{\rm CP}} c_{12} s_{13} s_{23} \right) 
 \\
&  + m_2 e^{2 i \alpha_2} \left(- c_{12} s_{23} - e^{i \delta_{\rm CP}} s_{12} s_{13} c_{23} \right) \left(c_{12} c_{23} - e^{i \delta_{\rm CP}} s_{12} s_{13} s_{23} \right) 
\\
&  + m_3 e^{-i \delta_{\rm CP}} c^2_{13} s_{23} c_{23}~, 
\\
(m^\nu)_{33} =&~
m_1 e^{2 i \alpha_1}  \left( s_{12} s_{23} - e^{i \delta_{\rm CP}} c_{12} s_{13} c_{23} \right)^2 
\\
&  + m_2 e^{2 i \alpha_2} \left( - c_{12} s_{23} - e^{i \delta_{\rm CP}} s_{12} s_{13} c_{23} \right)^2 
 + m_3   c^2_{13} c^2_{23}~. 
\end{split}
\label{eq:mnu_ij}
\end{align}

%%%%%%%%%%%%%%%%%%%%%%%%%%%%%%%%%%%%%%%%%%%%%%%%%%%%%%%%%%%%
\subsection{ Radiative lepton decay $\ell_i \to \ell_j \gamma$ } 
%%%%%%%%%%%%%%%%%%%%%%%%%%%%%%%%%%%%%%%%%%%%%%%%%%%%%%%%%%%%

Among the LFV processes, the most constraining process is the radiative muon decay $\mu\to e \gamma$.  It can be the most important process to discover LFV due to the high sensitivity to the new physics effects.  In order to study other radiative lepton decays, such as $\tau\to (e,\mu) \gamma$, in the following, we calculate the branching ratios for the $\ell_i \to \ell_j \gamma$ decays, where $\ell_j$ is the lighter lepton and its mass is neglected unless otherwise stated.

%%%%%%%%%%%%%%%%%%%%%%%%%%%%%%%%%%%%%%%%%%%%%%%%%%%%%%%%%%%%
\begin{figure}[phtb]
\begin{center}
\includegraphics[scale=1]{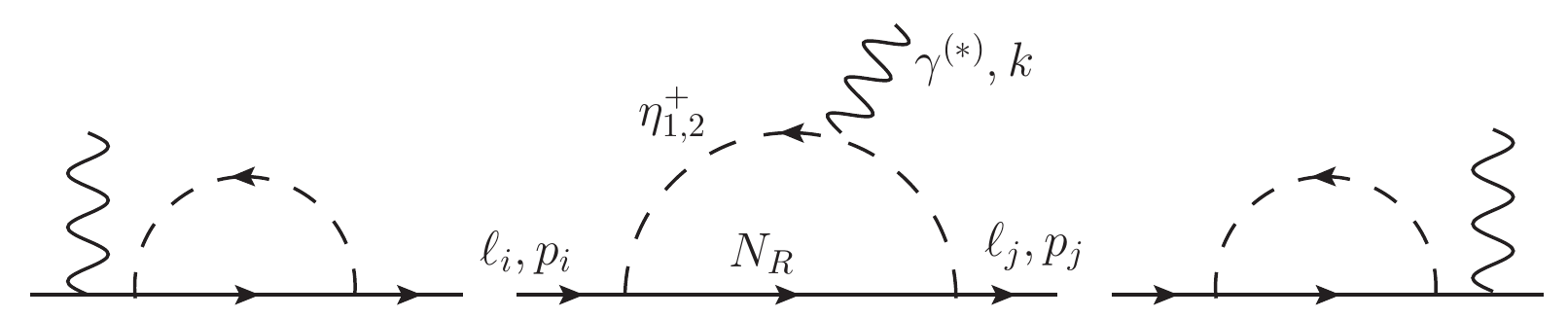}
\caption{One-loop Feynman diagrams for $\ell_i \to \ell_j \gamma^{(*)}$ }
\label{fig:photon_pen}
\end{center}
\end{figure}
%%%%%%%%%%%%%%%%%%%%%%%%%%%%%%%%%%%%%%%%%%%%%%%%%%%%%%%%%%%%

The radiative LFV process in the model arises from the $\eta^\pm_i$-mediated diagrams, as shown in Fig.~\ref{fig:photon_pen}.  Using the Yukawa couplings in Eq.~(\ref{eq:yuka}), the loop-induced effective Lagrangian for $\ell_i \to \ell_j \gamma^{(*)}$ can be expressed as:
\begin{equation}
{\cal L}^\gamma_{\ell_i \ell_j \gamma}  = e k^2 C^\gamma_{1 ji}  \, \overline\ell_j \gamma_\mu P_L \ell_i A^\mu - \frac{e m_{\ell_i}}{2}  C^\gamma_{2 ji} \, \overline\ell_j \sigma_{\mu \nu} P_R \ell_i F^{\mu\nu} ~, \label{eq:L_Cgamma}
\end{equation}
where the loop induced Wilson coefficients are obtained as:
 \begin{align}
 \begin{split}
 C^\gamma_{1ji} & = \sum^2_{\alpha=1} \frac{ y^*_{\alpha i} y_{\alpha j}}{(4\pi)^2 m^2_{\eta^\pm_\alpha} } J^\gamma_1\left( \frac{m^2_N}{m^2_{\eta^\pm_\alpha}}\right) ~, 
 \\
 C^\gamma_{2ji} & =  \sum^2_{\alpha=1}\frac{ y^*_{\alpha i } y_{\alpha j} }{(4\pi)^2 m^2_{\eta^\pm_\alpha}}  J^\gamma_{2} \left( \frac{m^2_N}{m^2_{\eta^\pm_\alpha}}\right)~, \label{eq:C_litoljga}
 \end{split}
\end{align}
and the loop integral functions are defined by
\begin{align}
\begin{split}
J^\gamma_1 (x) &= \frac{1}{36 (1-x)^4} \left(2-9 x + 18 x^2 -11 x^3 + 6 x^3 \ln x \right) ~,
\\
J^\gamma_2(x) & = \frac{1}{12(1-x)^4} \left( 1-6 x + 3 x^2+ 2 x^3 - 6 x^2 \ln x\right)~.
\end{split}
\end{align}
The emitted photon can be on-shell ($k^2=0$) or off-shell ($k^2\neq 0$), where the latter can be used to study the $\ell_i \to 3 \ell_j$ process, in which a $\ell_j^+ \ell_j^-$ pair is produced by the off-shell photon.  (The contribution of the $Z$-mediated diagrams is found to be small and will be neglected, as discussed in the next section.)  Using the effective Lagrangian in Eq.~(\ref{eq:L_Cgamma}), the branching ratio for $\ell_i \to \ell_j \gamma$ can be obtained as~\cite{Toma:2013zsa}:
 \begin{equation}
 BR(\ell_i \to \ell_j \gamma) = 
 \frac{3 (4\pi)^3 \alpha_{\rm em}}{4 G^2_F} \left| C^{\gamma}_{2ji} \right|^2 BR(\ell_i \to \ell_j \overline\nu_j \nu_i)~, 
 \label{eq:litoljga}
 \end{equation}
with $\alpha_{\rm em}=e^2/(4\pi)$.

In addition to the radiative LFV decays, the dipole operator in Eq.~(\ref{eq:L_Cgamma}) can contribute to the lepton anomalous  magnetic dipole moment ($g-2$) when we replace $\ell_j$ by $\ell_i$.  As a result, the lepton $(g-2)$ can be directly obtained as:
 \begin{equation}
a_\ell \equiv \left( \frac{g-2}{2} \right)_\ell = - m^2_\ell C^\gamma_{2\ell \ell}~. \label{eq:gm2}
 \end{equation}

%%%%%%%%%%%%%%%%%%%%%%%%%%%%%%%%%%%%%%%%%%%%%%%%%%%%%%%%%%%%
\subsection{$\ell_i \to 3 \ell_j $ from penguin and box diagrams} 
%%%%%%%%%%%%%%%%%%%%%%%%%%%%%%%%%%%%%%%%%%%%%%%%%%%%%%%%%%%%

We now discuss the three-body lepton flavor-changing decays.  For simplicity, we only concentrate on the $\ell_i \to 3 \ell_j$ decay.  In the model, the LFV processes $\ell_i \to 3 \ell_j$ can arise from photon-penguin, $Z$-penguin, and box diagrams.  In the following, we discuss their contributions in sequence.

Since the effective interaction for $\ell_i \to \ell_j \gamma^*$ has been given in Eq.~(\ref{eq:L_Cgamma}), the transition amplitude with the assigned momenta for the $\ell_i (p) \to \ell_j (p_1) \ell^+ (p_2) \ell^-(p_3)$ decay can be written as:
 \begin{align}
 \begin{split}
 {\cal M}_\gamma =&~ 
 e^2 C^\gamma_{1 ji}  \overline u(p_1) \gamma_\mu u(p)\, \overline u(p_3) \gamma^\mu v(p_2) 
 \\
 & +  \frac{e^2 m_{ \ell_i} }{k^2} C^{\gamma}_{2 ji} \overline u(p_1) i \sigma_{\mu\nu} k^\nu u(p)\, \overline u(p_3) \gamma^\mu v(p_2) 
 - ( p_1 \leftrightarrow p_3 )~,  
 \label{eq:ga_M}
 \end{split}
 \end{align}
where the photon coupling to the SM charged lepton shown in Eq.~(\ref{eq:Z}) is applied.

The Feynman diagrams for  $\ell_i \to \ell_j Z^*$ are similar to the photon diagrams shown in Fig.~\ref{fig:photon_pen}, where we use $Z^*$ instead of $\gamma^*$.  Using the $Z$ gauge couplings to $\eta^\pm$ and the SM lepton, given in Eqs.~(\ref{eq:gauge_kin}) and (\ref{eq:Z}), the one-loop induced effective interaction for $\ell_i\ell_j Z$ can be obtained as:
\begin{align}
{\cal L}^Z _{\ell_i \ell_j Z}&= \frac{ g c_{2W} }{ c_W } \sum^2_{\alpha=1}\frac{m_{\ell_i}  m_{\ell_j}}{(4\pi)^2 m^2_{\eta_\alpha}} J^\gamma_2\left( \frac{m^2_N}{m^2_{\eta_\alpha}}\right) \overline\ell_j \gamma_\mu P_R \ell_i Z^{\mu}~,
\end{align}
where we have dropped the small  contributions from $k^2$ and $k$ terms~\cite{Toma:2013zsa}.  It can be clearly seen that the loop-induced $Z$ coupling is proportional to the product of the lepton masses, {\it i.e.}, $m_{\ell_i} m_{\ell_j}$.  We note that because $N$ does not couple to $Z$, the $Z$ boson cannot be emitted from the fermion line in the loop; therefore, no chiral enhancement factor, {\it e.g.}, $m^2_N/m^2_{\eta^\pm_i}$, appears in the model.  When a necessary mass insertion occurs in the initial lepton $\ell_i$, in order to balance the chirality of the lepton vector current that couples to the vector gauge boson $Z$, a mass factor is necessarily inserted in the final lepton $\ell_j$.  Hence, we get the result proportional to $m_{\ell_i} m_{\ell_j}$.   Due to the fact that $m_{\ell_j} \ll m_{\ell_i}$ and  $m_{\ell_i} m_{\ell_j}/m^2_Z \ll 1$, we thus neglect the $Z$-penguin contributions to $\ell_i \to 3 \ell_j$.

%%%%%%%%%%%%%%%%%%%%%%%%%%%%%%%%%%%%%%%%%%%%%%%%%%%%%%%%%%%%
\begin{figure}[phtb]
\begin{center}
\includegraphics[scale=1]{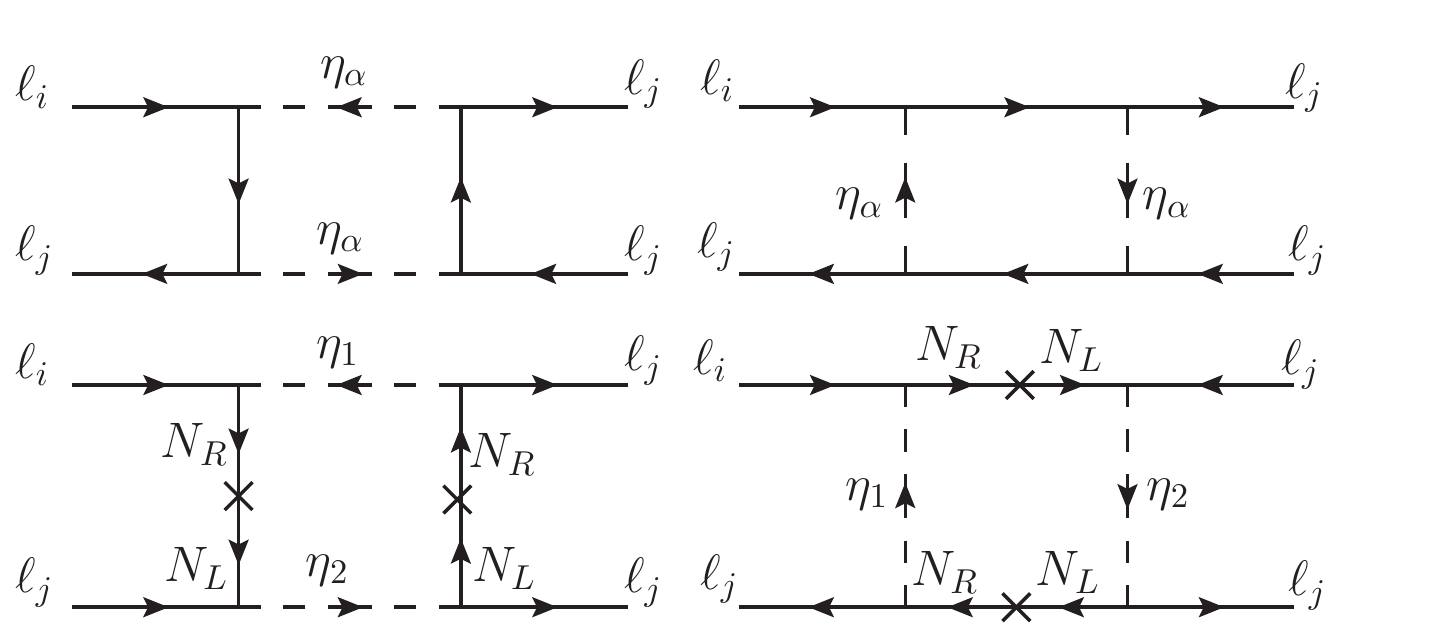}
\caption{Selected box diagrams for $\ell_i \to 3\ell_j $ }
\label{fig:boxes}
\end{center}
\end{figure}
%%%%%%%%%%%%%%%%%%%%%%%%%%%%%%%%%%%%%%%%%%%%%%%%%%%%%%%%%%%%

The box diagrams contributing to $\ell_i \to 3 \ell_j$ are partly shown in Fig.~\ref{fig:boxes}.  In addition to the diagrams mediated by the same $\eta^\pm_{1(2)}$, the other possible diagrams are mediated by both $\eta^\pm_1$ and $\eta^\pm_2$ and involve chirality-flipping factor $m^2_N$.  Using the Yukawa couplings in Eq.~(\ref{eq:yuka}), the four-fermion interaction for $\ell_i \to 3 \ell_j$ is obtained as:
\begin{equation}
{\cal M}_{\rm box} =  ( C^{\rm box}_{1ji} + C^{\rm box}_{2 ji} )  \, \overline\ell_j \gamma_\mu P_L \ell_j \,\overline\ell_j \gamma^\mu P_L \ell_i ~, 
\end{equation}
where $C^{\rm box}_1$ denotes the contributions from diagrams involving the same $\eta^\pm_{1(2)}$ and $C^{\rm box}_{2}$ are from those involving both $\eta^\pm_1$ and $\eta^\pm_2$, and the effective Wilson coefficients are given by:
 \begin{align}
 \begin{split}
 C^{\rm box}_{1ji} & = -\sum^2_{\alpha=1} \frac{ y^*_{\alpha i} y_{\alpha j} |y_{\alpha j}|^2}{ (4\pi )^2m^2_{\eta_\alpha}} J^{\rm box}_1\left( \frac{m^2_N}{m^2_{\eta_\alpha} }\right) ~, 
 \\
 C^{\rm box}_{2ji} & = \frac{ y^*_{1 i} y_{1 j} |y_{2 j}|^2 + y^*_{2i} y_{2 j} |y_{1j}|^2}{ (4\pi)^2 m^2_{\eta_2}}  \frac{m^2_N}{m^2_{\eta_2}} J^{\rm box}_2 \left( \frac{m^2_N}{ m^2_{\eta_2} }, \frac{m^2_{\eta_1}}{m^2_{\eta_2} }\right)~,
 \end{split}
\end{align}
with the loop integral functions given by
\begin{align}
\begin{split}
J^{\rm box}_1(x) & = \frac{1+x}{2(1-x)^2} + \frac{\ln x}{(1-x)^3} ~, 
\\
J^{\rm box}_2(x,y) & = - \frac{1}{(1-x) (y-x)} - \frac{(y-x^2) \ln x}{(1-x)^2 (y-x)^2} + \frac{y\ln y}{(1-y) (y-x)^2}~.
\end{split}
\end{align}
Although $C^{\rm box}_{2 ji}$ seems to have a chiral enhancement, its contribution is in fact somewhat smaller than $C^{\rm box}_{1 ji}$ due to the assumed condition that $m_N < m_{\eta_2}$.

Combining the contributions from the photon-penguin and box diagrams, the branching ratio for the $\ell_i \to 3 \ell_j$ decay can be written as~\cite{Toma:2013zsa}:
 \begin{align}
 \begin{split}
 BR(\ell_i \to 3 \ell_j) =&~ \frac{3 (4 \pi \alpha_{\rm em})^2}{8 G^2_F} \Bigg[
  |C^\gamma_{1ji}|^2 + |C^\gamma_{2 ji} |^2 \left( \frac{16}{3} \ln\left( \frac{m_i}{m_j} \right) - \frac{22}{3} \right)+ \frac{|C^{\rm box}_{ji} |^2}{6 (4\pi\alpha_{\rm em})^2}  
  \\
  &  + \frac{2}{3}\left( -6 \operatorname{Re}(C^\gamma_{1ji} C^{\gamma*}_{2ji}) 
  + \frac{1}{(4\pi\alpha_{\rm em})^2} \operatorname{Re}\left((C^\gamma_{1 ji}-2 C^\gamma_{2 ji}) C^{\rm box *}_{ji} \right)\right) \Bigg]
  \\
  & \times BR(\ell_i \to \ell_j \overline\nu_j \nu_i)~, \label{eq:3ell}
 \end{split}
\end{align}
where $C^{\rm box}_{ji} = C^{\rm box}_{1ji} + C^{\rm box}_{2ji}$.

%%%%%%%%%%%%%%%%%%%%%%%%%%%%%%%%%%%%%%%%%%%%%%%%%%%%%%%%%%%%
\subsection{$\mu-e$ conversion in nuclei and  muonium-antimuonium oscillation}
%%%%%%%%%%%%%%%%%%%%%%%%%%%%%%%%%%%%%%%%%%%%%%%%%%%%%%%%%%%%

The $\mu-e$ conversion process describes the muon capture by a nucleus through the process $\mu (A,Z) \to e (A,Z)$. At the quark level, the process can be represented as $\mu q \to e q$, as induced by the photon- and $Z$-penguin diagrams in the model.   Following the results in~\cite{Kuno:1999jp,Arganda:2007jw}, the conversion rate, which is relative to the muon capture rate, is given by:
\begin{align}
\begin{split}
CR(\mu-e,{\rm Nucleus}) =&~ 
\frac{p_e E_e m^3_\mu G^2_F \alpha^3_{\rm em} Z^4_{\rm eff} F^2_p}{8 \pi^2 Z \Gamma_{\rm cap}}
\\ 
& \times  \left[ \left| A_{+}  \left( g^{(0)}_{LV} + g^{(0)}_{LS} \right) + A_{-}  \left( g^{(1)}_{LV} + g^{(1)}_{LS} \right) \right|^2 + \left( R\leftrightarrow L \right)\right]~, 
\label{eq:CRmue}
\end{split}
\end{align}
where $p_e~(E_e)$ is the momentum (energy) of electron and is taken to be $m_{\mu}$ in the numerical estimates, $Z_{\rm eff}$ denotes the effective atomic charge of the nucleus, $F_p$ is the nuclear matrix element, $\Gamma_{\rm cap}$ is the total muon capture rate, $A_\pm = Z\pm \tilde{N}$ with $Z~(\tilde{N})$ being the proton (neutron) number in the nucleus, and $g^{(j)}_{\chi K}$ with $\chi=L,R$ and $K=V,S$ are defined by
 \begin{align}
\begin{split}
 g^{(0)}_{\chi K} & = \frac{1}{2} \sum_{q=u,d,s} \left( g^q_{\chi K} G^{(q,p)}_K + g^q_{\chi K} G^{(q,n)}_K\right)~, 
 \\
 g^{(1)}_{\chi K} & = \frac{1}{2} \sum_{q=u,d,s} \left( g^q_{\chi K} G^{(q,p)}_K - g^q_{\chi K} G^{(q,n)}_K\right)~. 
\end{split}
 \end{align}
Since the $Z$-penguin contribution is negligible, the dominant effect is from the photon-penguin diagrams.  Thus, the only nonzero $g^q_{\chi K}$ is:
 \begin{equation}
 g^q_{LV}= \frac{\sqrt{2} e^2 Q_q}{G_F} (C^\gamma_{1e\mu} - C^\gamma_{2 e\mu})~. \label{eq:gqLV}
 \end{equation}
The nucleon matrix element $G^{(q,N')}_K$ is  defined by $\langle N' | \overline q \Gamma_K q | N' \rangle = G^{(q,N')}_K \overline N' \Gamma_K N'$, where $\Gamma_{K}=(1, \gamma_\mu)$ when $K=(S,V)$, respectively. Their values can be found in Refs.~\cite{Kuno:1999jp,Kosmas:2001mv} as:
 \begin{align}
 G^{(u,p)}_V = G^{(d,n)}_V=2~, ~~G^{(d,p)}_V =G^{(u,n)}_V=1~, ~~ G^{(s,N')}=0~.
 \end{align}
For heavy nuclei, because $A_{-} \ll A_{+}$ and $g^{(1)}_{\chi K} < g^{(0)}_{\chi K}$, the $\mu-e$ conversion rate in the model can be simplified as:
 \begin{equation}
CR(\mu-e,{\rm Nucleus})  \approx  \frac{p_e E_e m^3_\mu G^2_F \alpha^3_{\rm em} Z^4_{\rm eff} F^2_p}{8 \pi^2 Z \Gamma_{\rm cap}} A^2_{+} \left|   g^{(0)}_{LV}  \right|^2 ~.
 \end{equation}

In addition to the $\mu-e$ conversion and LFV processes, which are $\Delta L_\mu=1$ processes, another interesting process involving $\Delta L_\mu=2$ ($\Delta L=2$ will be used hereinafter) is the muonium-antimuonium oscillation, where the muonium is a bound state of $\mu^+$ and $e^-$, {\it i.e.}, $|M_{\mu}\rangle \equiv | \mu^+ e^-\rangle$.  Similar to the $\mu\to 3e$ decay, the $M_\mu$-$\overline M_\mu$ mixing matrix element can arise from the box diagrams, where the Feynman diagrams are similar to Fig.~\ref{fig:boxes} with $(\ell_i,\ell_j) = (\mu,e)$.  As in the case of meson oscillations, $M_\mu$-$\overline M_\mu$ mixing effect can be taken as a perturbative effect in quantum mechanics and is formulated by~\cite{Conlin:2020veq}
 \begin{equation}
 \left(m - \frac{i}{2} \Gamma \right)_{12} = \frac{1}{2m_{M_\mu}} \langle \overline M_\mu| {\cal H}_{\rm eff} | M_\mu \rangle + \frac{1}{m_{M_\mu}} \sum_n \frac{\langle \overline M_\mu | {\cal H}_{\rm eff} | n \rangle \langle n | {\cal H}_{\rm eff} | M_\mu \rangle}{m_{M_\mu} -E_n + i\epsilon}~, \label{eq:DL=2}
 \end{equation}
where $m_{12}$ and $\Gamma_{12}$ lead to the mass and lifetime differences between the two physical states of muonium, and  $n$ in the second term denotes the possible intermediate state.  In order to produce the lifetime difference $\Delta \Gamma$, we need a resonant intermediate state.  In the model, the new particle masses are heavier than the muonium mass $m_{M_\mu}$.  As such, $\Delta \Gamma$ by the new effects is negligible, and we concentrate on the mass difference $\Delta m=m_1 - m_2\approx 2 \operatorname{Re} (m_{12})$. From Eq.~(\ref{eq:DL=2}), the parameter $x\equiv \Delta m/\Gamma_\mu$ used to show the probability of $M_\mu\to  \overline M_\mu$ can be written as:
 \begin{equation}
 x\approx \frac{1}{m_{M_\mu} \Gamma_\mu} \operatorname{Re}(\langle \overline M_\mu| {\cal H}_{\rm eff} | M_\mu \rangle)~.
 \end{equation}

Following the formulation obtained in Ref.~\cite{Conlin:2020veq}, the oscillation probability is:
  \begin{equation}
  P(M_\mu \to \overline M_\mu) = \frac{1}{2} (x^2 + y^2) \approx \frac{x^2}{2}~,
  \end{equation}
while the experimental upper limit is:
  \begin{equation}
  P(M_\mu \to \overline M_\mu)^{\rm exp} \leq 8.3 \times 10^{-11}/S_B(B_0)~,
  \end{equation}
  with $S_B(B_0)=0.75$~\cite{Willmann:1998gd,Conlin:2020veq} taken in this work.  Since the spin-0 para-muonium and spin-1 ortho-muonium are produced in the experiment~\cite{Willmann:1998gd}, in order to compare the theoretical estimate with the experimental data, we follow the prescription given in Ref.~\cite{Conlin:2020veq} and take the spin average by combing both spin-0 and spin-1 muonia as:
  \begin{equation}
  P(M_\mu \to \overline M_\mu)^{\rm exp} = \sum_{i=P,V}  \frac{1}{2 s_{i} + 1} P(M^i_\mu \to \overline M^i_\mu)~, \label{eq:muonium_P}
  \end{equation}
  where $s_i$ denotes the spin of the muonium $M^i_\mu$.

According to the Yukawa couplings shown in Eq.~(\ref{eq:yuka}) and the Feynman diagrams in Fig.~\ref{fig:boxes}, the effective interaction for the  $\Delta L=2$ process can be written  as:
\begin{equation}
{\cal H}_{\rm eff} = C^{\Delta L=2}_1 \left( \overline \mu \gamma_\mu P_L e \right) \left( \overline \mu \gamma^\mu P_L e \right)~, 
\end{equation}
where the Wilson coefficient $C^{\Delta L=2}_1$ is given by:
 \begin{equation}
 C^{\Delta L=2}_1  =\sum^2_{\alpha=1} \frac{ (y_{\alpha \mu} y^*_{\alpha e})^2}{ (4\pi )^2m^2_{\eta_\alpha}} J^{\rm box}_1\left( \frac{m^2_N}{m^2_{\eta_\alpha} }\right) 
 - 2 \frac{ y_{1 \mu} y^*_{1 e}  y_{2 \mu} y^*_{2 e} }{ (4\pi)^2 m^2_{\eta_2}}  \frac{m^2_N}{m^2_{\eta_2}} J^{\rm box}_2 \left( \frac{m^2_N}{ m^2_{\eta_2} }, \frac{m^2_{\eta_1}}{m^2_{\eta_2} }\right)~.
 \end{equation}
Using the transition matrix element $\langle \overline M_\mu | O_1 | M_\mu \rangle = f^2_{M_\mu} m^2_{M_\mu}$, the $x$-parameter for para-muonium and ortho-muonium are:
 \begin{align}
 x_P &= \frac{4 (m_{\rm red} \alpha_{\rm em} )^3}{\pi \Gamma_\mu} C^{\Delta L=2}_1~, \nonumber \\
 x_V & = - \frac{12 (m_{\rm red} \alpha_{\rm em})^3}{\pi \Gamma_\mu} C^{\Delta L=2}_1~,
 \end{align}
 where the reduced mass $m_{\rm red} = m_\mu m_e/(m_\mu + m_e)$, and $f^2_{M_\mu} = 4(m_{\rm red} \alpha_{\rm em})^3/m_{M_\mu}$~\cite{Conlin:2020veq}. Hence, Eq.~(\ref{eq:muonium_P}) can be expressed as:
 \begin{equation}
 P(M_\mu \to \overline M_\mu)^{\rm exp} = \frac{x^2_P}{2} + \frac{x^2_V}{6}~. \label{eq:P_muonium}
 \end{equation}

%%%%%%%%%%%%%%%%%%%%%%%%%%%%%%%%%%%%%%%%%%%%%%%%%%%%%%%%%%%%
\subsection{Loop-induced kinetic mixing between $U(1)_Y$ and $U(1)_X$}
%%%%%%%%%%%%%%%%%%%%%%%%%%%%%%%%%%%%%%%%%%%%%%%%%%%%%%%%%%%%

%%%%%%%%%%%%%%%%%%%%%%%%%%%%%%%%%%%%%%%%%%%%%%%%%%%%%%%%%%%%
\begin{figure}[phtb]
\begin{center}
\includegraphics[scale=0.5]{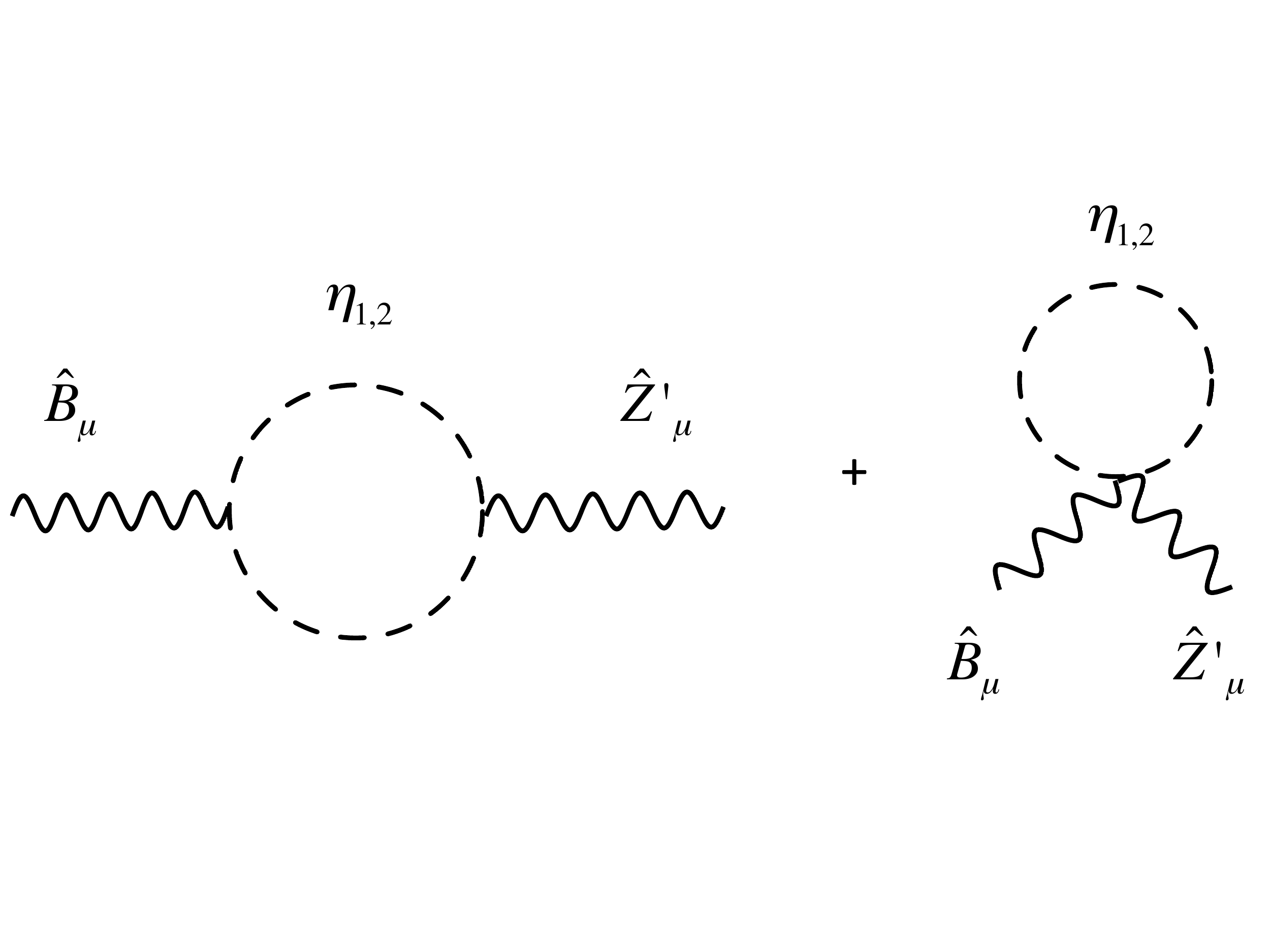}
\caption{One-loop diagrams inducing kinetic mixing between $U(1)_Y$ and $U(1)_X$ gauge fields. }
\label{fig:kinetic-mixing}
\end{center}
\end{figure}
%%%%%%%%%%%%%%%%%%%%%%%%%%%%%%%%%%%%%%%%%%%%%%%%%%%%%%%%%%%%

In our model the SM particles do not interact with $Z'$ boson at tree level since they are not charged under $U(1)_X$.
However they interact with $Z'$ through kinetic mixing between $U(1)_Y$ and $U(1)_X$ gauge fields induced by the one-loop diagrams shown in Fig.~\ref{fig:kinetic-mixing}.
We calculate these diagrams and obtain kinetic mixing term 
\begin{align}
%\begin{split}
 \mathcal{L}_{KM} &= - \frac{\sin \epsilon}{2} \hat{B}_{\mu \nu} \hat{Z'}^{\mu \nu} ~, 
\\
 \sin \epsilon & =  \frac{Q_X g_X g'}{2 (4 \pi^2 ) } \int_0^1 dx (1-x)(2x-1)^2  \nonumber \\
& \times\left( \ln \left[ \frac{x(1-x) q^2 - m_{\eta^\pm_2}^2}{x(1-x) q^2 - m_{\eta^\pm_1}^2} \right] + \ln \left[ \frac{x(1-x) q^2 - m_{S_2}^2}{x(1-x) q^2 - m_{S_1}^2} \right] \right) ~,
%\end{split}
\end{align}
where $q^2$ corresponds to the momentum carried by the gauge bosons.
Note that the diagrams sum up to give a finite result for the kinetic mixing with divergences canceled between the contributions from the two inert doublet scalars carrying opposite charges under $U(1)_X$.
For $q^2 \ll m^2_{\eta_{1,2}^\pm, S_{1,2}}$, we can approximate the formula of $\epsilon$ by
\begin{equation}
\sin \epsilon \sim \frac{Q_X g_X g'}{3 (4 \pi)^2} \left( \ln \frac{m_{\eta^\pm_2}}{m_{\eta^\pm_1}} + \ln \frac{m_{S_1}}{m_{S_2}} \right).
\end{equation}
We can diagonalize the kinetic terms for $\hat{Z'}_\mu$ and $B_\mu$ via the following transformations~\cite{Babu:1997st}:
\begin{align}
\begin{split}
& \hat{B}_\mu = B_\mu - \tan { \epsilon} Z'_\mu ~, 
\\
& \hat{Z'_\mu} = \frac{1}{\cos { \epsilon} } Z'_\mu ~.
\end{split}
\end{align}
where we approximate $\sin \epsilon \simeq \epsilon$ and $\cos \epsilon \simeq 1$ hereafter.
After electroweak symmetry breaking, we obtain the mass terms for neutral gauge bosons as
\begin{equation}
\mathcal{L}_M = \frac12 m_{ Z}^2 \tilde Z_\mu \tilde Z^\mu + m_{Z}^2 \epsilon s_W Z'_\mu \tilde{Z}^\mu + \frac12 m_{Z'}^2 Z'_\mu Z'^\mu
~,
\end{equation}
where $m_{Z} = v \sqrt{g^2+g'^2}/2$ and $\tilde{Z}_\mu = c_W W^3_\mu - s_W B_\mu$.
We then obtain mass eigenstates and eigenvalues of the neutral gauge bosons as 
\begin{align}
\begin{split}
& \{ m_{Z_1}^2, m_{Z_2}^2 \} = 
\frac{1}{2} (m_{Z'}^2 + m_{Z}^2 ) \mp  \frac{1}{2} \sqrt{(m_{Z'}^2- m_{Z}^2)^2 + 4 \epsilon^2 s_W^2 m_{Z}^4}
~, \\
&
\begin{pmatrix} Z_{1\mu} \\ Z_{2\mu} \end{pmatrix} = 
\begin{pmatrix} \cos \theta_{ZZ'} & -\sin \theta_{ZZ'} \\ \sin \theta_{ZZ'} & \cos \theta_{ZZ'} \end{pmatrix}
\begin{pmatrix} \tilde{Z}_\mu \\ Z'_\mu \end{pmatrix} ~, \quad
  \tan 2 \theta_{ZZ'} =  \frac{2 s_W \epsilon m_{Z}^2}{m_{Z}^2 - m_{Z'}^2}
  ~,
\end{split}
\end{align}
where we can approximate $m_{Z_1} \simeq m_{Z}$ and $m_{Z_2} \simeq m_{Z'}$ for tiny $\epsilon$.
The $Z_1$ and $Z_2$ bosons are to be identified as the physical massive gauge bosons and, to avoid the pedantry while being generally not confusing, will be referred to as the $Z$ and $Z'$ bosons.  The $Z'$ interaction with SM fermions $f$ is now given by
\begin{align}
\label{eq:Zpff}
\mathcal{L}_{Z' \overline f f} & = \sum_{\chi= L,R} \frac{g}{\cos \theta_W} Z'_\mu \overline f_\chi \gamma^\mu \left[  S_{ZZ'} (T_3 - Q \sin^2 \theta_W) + C_{ZZ'} \epsilon Y  \sin \theta_W  \right]   f_\chi \nonumber \\
& \equiv  Z'_\mu \overline f \gamma^\mu (C_L^f P_L + C_R^f P_R) f
~,
\end{align} 
where $T_3$ is diagonal generator of $SU(2)_L$, $Q$ is the electric charge operator, $S_{ZZ'} \equiv \sin \theta_{ZZ'}$, and $C_{ZZ'} \equiv \cos \theta_{ZZ'}$.

%%%%%%%%%%%%%%%%%%%%%%%%%%%%%%%%%%%%%%%%%%%%%%%%%%%%%%%%%%%%
\section{Numerical Analysis} \label{sec:NA}
%%%%%%%%%%%%%%%%%%%%%%%%%%%%%%%%%%%%%%%%%%%%%%%%%%%%%%%%%%%%

In this section, we perform numerical scans for the allowed parameter space and the corresponding ranges of various observables, which are then compared with current experimental bounds or the sensitivities that ongoing/future experiments can probe.  We also calculate the DM relic density and check against the constraint of the direct search limit.

%%%%%%%%%%%%%%%%%%%%%%%%%%%%%%%%%%%%%%%%%%%%%%%%%%%%%%%%%%%%
\subsection{Inputs and constraints}
%%%%%%%%%%%%%%%%%%%%%%%%%%%%%%%%%%%%%%%%%%%%%%%%%%%%%%%%%%%%

From Eq.~(\ref{eq:mnu_ij}), it can be seen that when the neutrino mixing angles and masses are determined, the parameters of ${\bf y}_{1,2}$, $m_{S_i, A_i}$, and $m_N$ can be bounded. In order to get the allowed ranges for the free parameters, we take the values of the neutrino oscillation parameters obtained from the global fit in Refs.~\cite{Esteban:2020cvm, Gonzalez-Garcia:2021dve} as the inputs, where the global fit results with $3\sigma$ errors are given in Table~\ref{tab:nu_GF}.  Based upon the fit results, the ranges of the Majorana mass matrix elements in units of eV for the normal ordering (NO) and inverted ordering (IO) can be respectively estimated as:
\begin{align}
\begin{split}
\begin{pmatrix} |m^\nu_{11}| & |m^\nu_{12}| & |m^\nu_{13}| \\ |m^\nu_{21}| & |m^\nu_{22}| & |m^\nu_{23}| \\ |m^\nu_{31}| & |m^\nu_{32}| & |m^\nu_{33}| \end{pmatrix}_{\rm NO} & 
\simeq \begin{pmatrix} 0.09-0.42 & 0.095-0.909 & 0.087-0.906 \\  0.095-0.909 & 1.51-3.31 & 2.03-2.81 \\ 0.087-0.906 & 2.03-2.81  & 1.46-3.27\end{pmatrix} \times 10^{-2}~, 
\\
\begin{pmatrix} |m^\nu_{11}| & |m^\nu_{12}| & |m^\nu_{13}| \\ |m^\nu_{21}| & |m^\nu_{22}| & |m^\nu_{23}| \\ |m^\nu_{31}| & |m^\nu_{32}| & |m^\nu_{33}| \end{pmatrix}_{\rm IO} & 
\simeq \begin{pmatrix} 1.50-4.97 & 0.-3.81 & 0.-3.88 \\  0.-3.81 & 0.-3.05 & 0.65-2.60 \\ 0.-3.88 & 0.65-2.60 & 0.-3.16\end{pmatrix} \times 10^{-2}
~, 
\end{split}
\label{eq:v_nu_mass}
\end{align}
 where  $m_{1(3)}=0$ for NO (IO) is applied. Since we do not have any information on the Majorana phases at the moment, we take $\alpha_{1,2} \in [-\pi, \pi]$. We then use Eq.~(\ref{eq:v_nu_mass}) to bound the free parameters.

%%%%%%%%%%%%%%%%%%%%%%%%%%%%%%%%%%%%%%%%%%%%%%%%%%%%%%%%%%%%
\begin{table}[htp]
\caption{Global fit with $3\sigma$ ranges based upon the observations of neutrino oscillations~\cite{Esteban:2020cvm,Gonzalez-Garcia:2021dve}.}
\begin{center}
\begin{tabular}{c|cccccc} \hline \hline
 ~~& $\theta_{12}/^\circ$ & $\theta_{13}/^\circ$ & $\theta_{23}/^\circ$ & $\delta_{\rm CP}/^\circ$ & $\Delta m^2_{21} \times 10^{5}$ [eV$^2$] ~& ~$\Delta m^2_{3\ell} \times 10^{3}$ [eV$^2$]   \\ \hline
 NO & ~$(31.27, 35.87)$~& ~$(8.25, 8.98)$~ & ~$(39.7,50.9)$~ & ~$(144, 350)$~ & ~$(6.82,  8.04)$~ & ~ $(2.43, 2.59)$ \\ \hline 
 IO  & $(31.27, 35.87)$ & $(8.24,9.02)$ & $(39.8, 51.6)$ & $(194, 345)$ & $(6.82, 8.04)$ & $(-2.574, -2.410)$  \\ \hline \hline

\end{tabular}
\end{center}
\label{tab:nu_GF}
\end{table}%
%%%%%%%%%%%%%%%%%%%%%%%%%%%%%%%%%%%%%%%%%%%%%%%%%%%%%%%%%%%%

The parameter space can be further constrained by various LFV processes, whose current upper bounds are given in Table~\ref{tab:upper_exp}. With $S_B(B_0)=0.75$, the upper limit on the probability of the muonium oscillation is taken as $P(M_\mu \to \overline M_\mu)^{\rm exp} <  11.1 \times 10^{-11}$~\cite{Willmann:1998gd,Conlin:2020veq}.

%%%%%%%%%%%%%%%%%%%%%%%%%%%%%%%%%%%%%%%%%%%%%%%%%%%%%%%%%%%%
\begin{table}[htp]
\caption{Current upper bounds on various LFV processes.  Except the $\mu-e$ conversion rate in titanium quoted from Ref.~\cite{SINDRUMII:1993gxf}, all other values are obtained from the Particle Data Group~\cite{PDG}.}
\begin{center}
\begin{tabular}{c|cccccccc} \hline \hline
 ~~& $\mu \to e \gamma$ & $\mu \to 3 e$ & $\tau\to e \gamma$ & $\tau\to \mu \gamma$ & $\tau\to 3e$ & $\tau\to 3\mu$ & $CR(\mu-e,{\rm Ti})$      \\ \hline 
 U.L.  &  $4.2 \times 10^{-13}$  & $1.0 \times 10^{-12}$ &  $3.3\times 10^{-8}$ & $4.4\times 10^{-8}$  & $2.7\times 10^{-8}$ & $2.1\times 10^{-8}$& $4.3 \times 10^{-12}$~\cite{SINDRUMII:1993gxf} \\ \hline \hline
\end{tabular}
\end{center}
\label{tab:upper_exp}
\end{table}%
%%%%%%%%%%%%%%%%%%%%%%%%%%%%%%%%%%%%%%%%%%%%%%%%%%%%%%%%%%%%

There are $12$ free parameters involved in the model, and we only have $6$ observables from the neutrino oscillation experiments. In order to  make the parameter scans more efficient, following the strict constraint from the $\mu\to e\gamma$ decay shown in Eqs.~(\ref{eq:C_litoljga}) and (\ref{eq:litoljga}), we assume
\begin{equation}
 y_{2 \mu} = - \frac{m^2_{S_2}}{m^2_{S_1}} \frac{y^{*}_{1e}}{y^*_{2e}} \frac{J^\gamma_2\left( \frac{m^2_{N}}{m^2_{S_1}}\right)}{J^\gamma_2\left( \frac{m^2_{N}}{m^2_{S_2}}\right)} + \xi e^{i \phi}~,
\end{equation}
where $\xi$ is a real parameter and $\phi$ is its phase.  The parameter ranges used to scan the parameter space are taken as follows: 
\begin{align}
\begin{split}
  & m_N \in [100, 1000]~{\rm GeV}~,~m_{S_{1,2}}\in [800, 2500]~ {\rm GeV}~, 
  \\
  & y_{1i, 2i\neq \mu} \in [-1,1]~,~\phi_{1,3} \in [-\pi, \pi]~,~\xi\in [-0.1,0.1]~,~ \phi\in [-\pi, \pi]~.
\end{split}
\end{align}
In addition, we also assume $m_N < m_{S_1} < m_{S_2}$ and $m_{\eta^\pm_{i}} = m_{S_{i}}$ in the numerical estimates.

\begin{figure}[phtb]
\begin{center}
\includegraphics[scale=0.6]{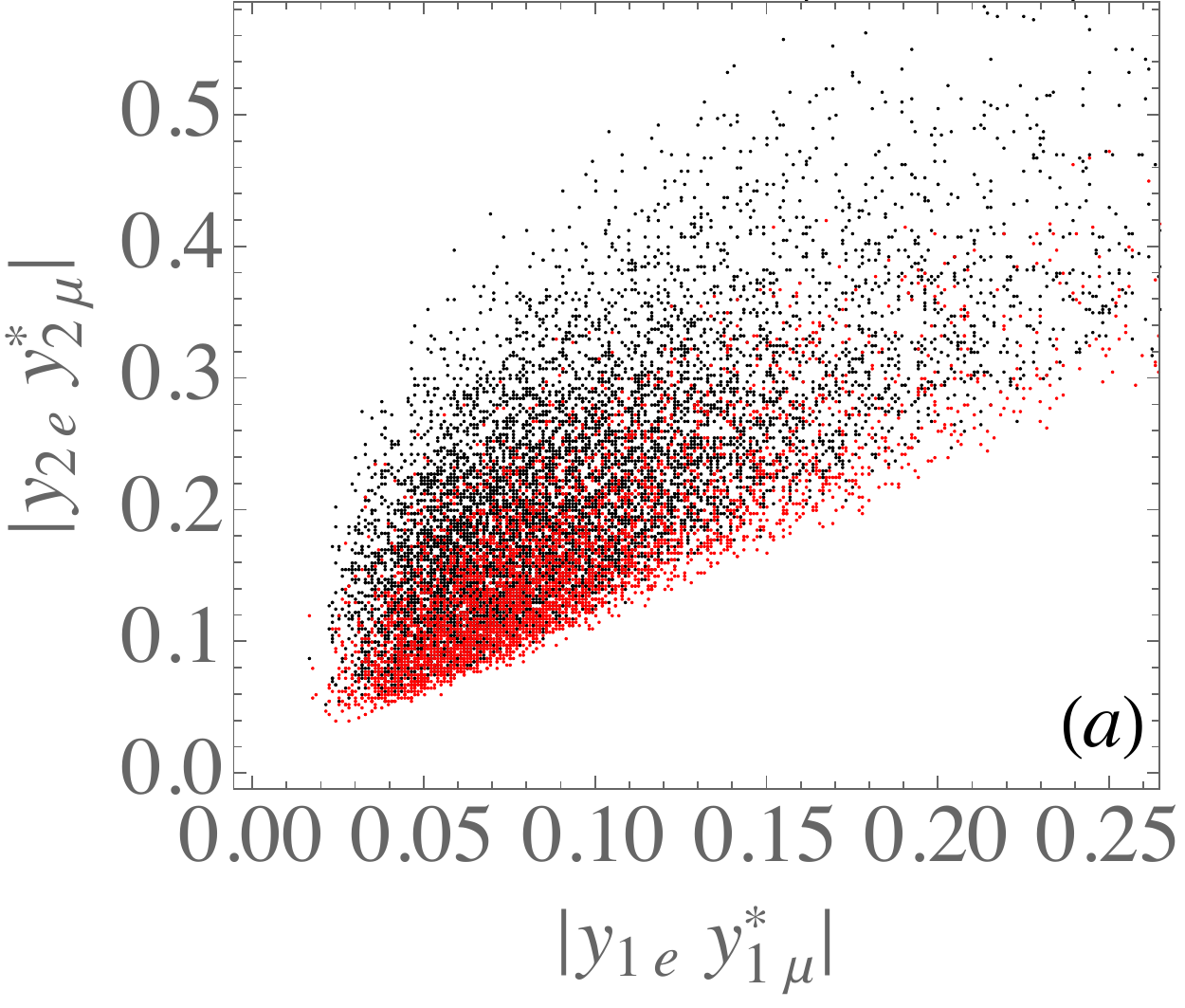}
\includegraphics[scale=0.6]{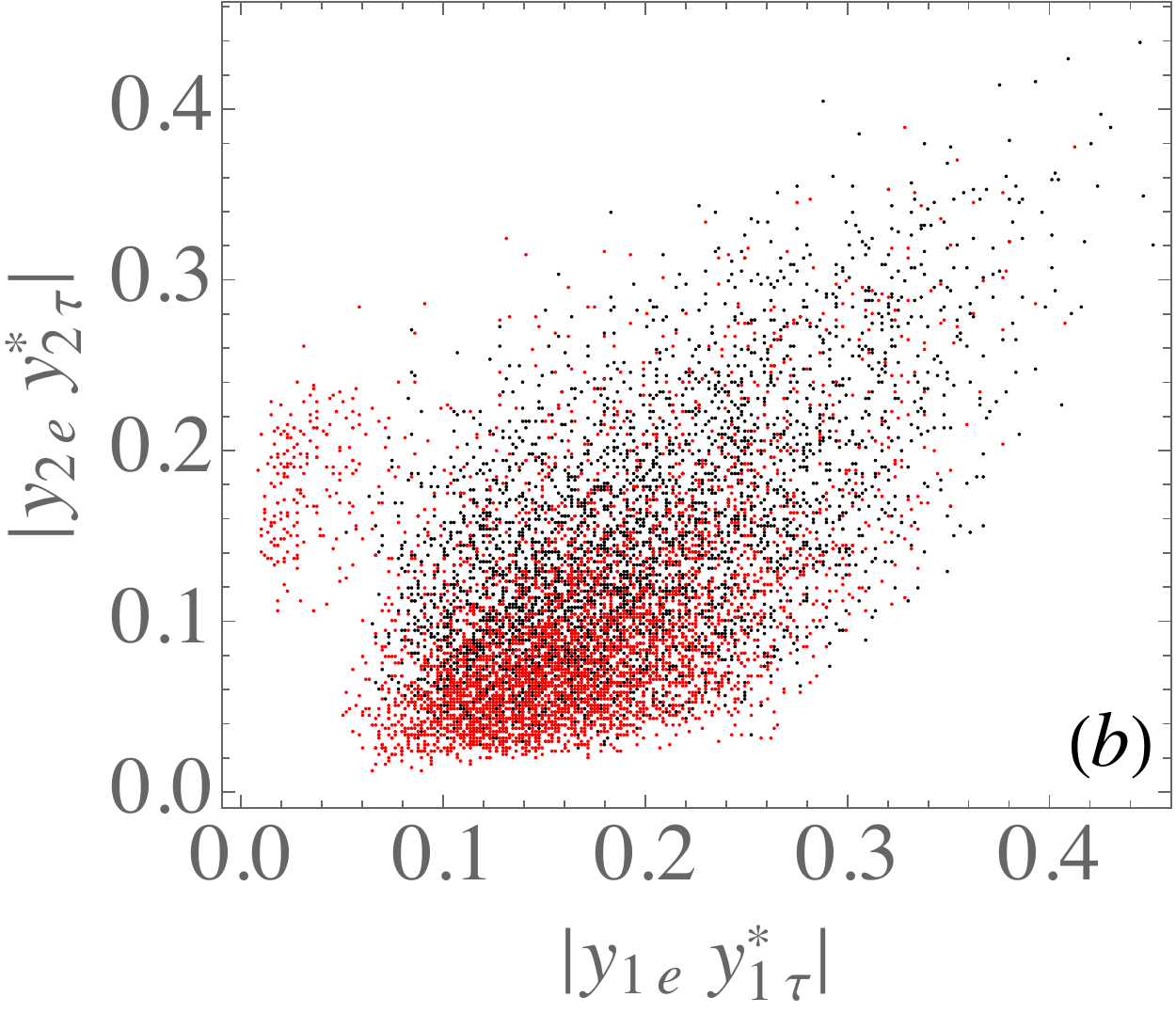}
\includegraphics[scale=0.6]{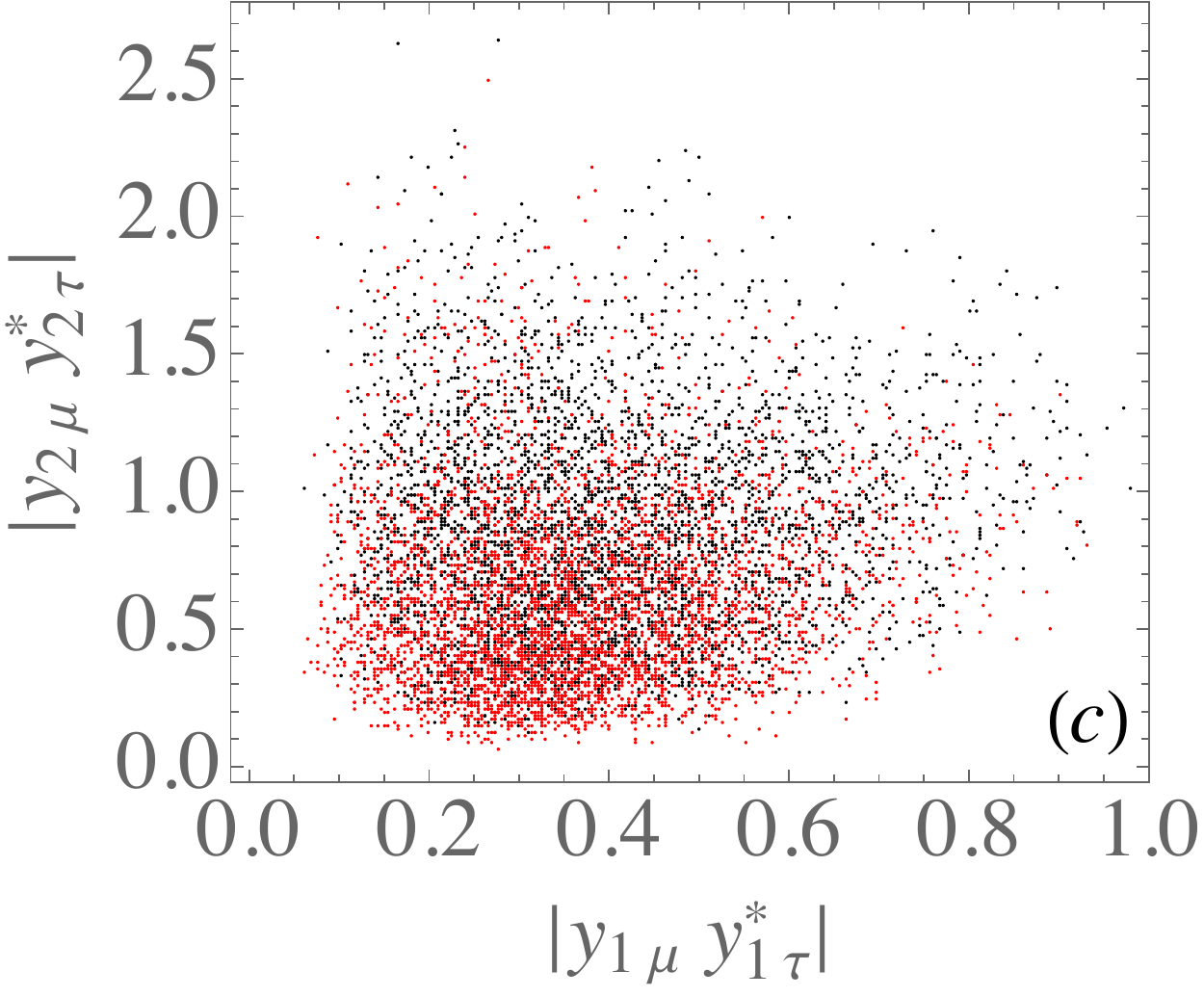}
\includegraphics[scale=0.6]{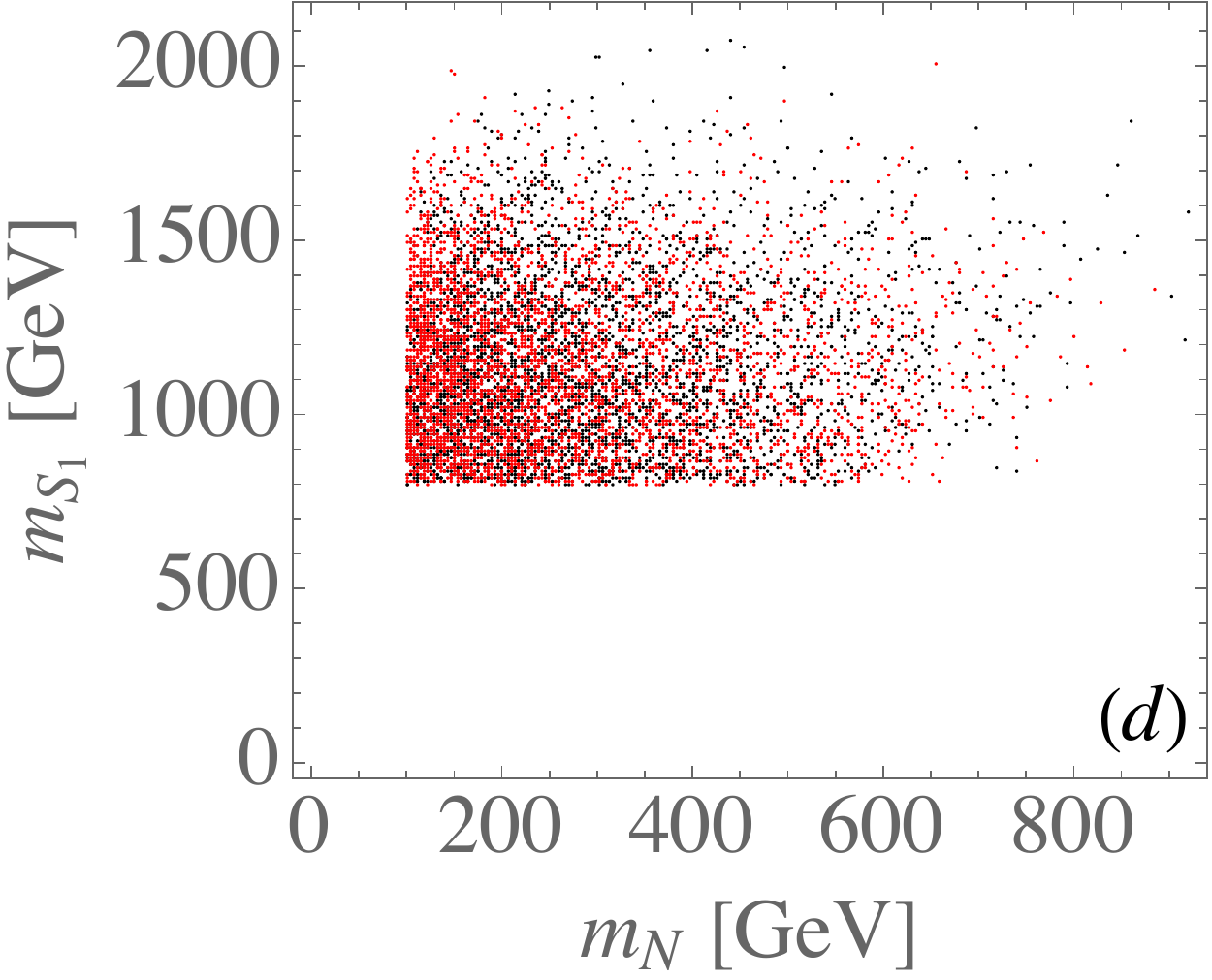}
\caption{ (a)-(c) Constraints on various products of Yukawa couplings, and (d) allowed ranges of $m_{N,S_1}$.  The black and red points denote the results of NO and IO, respectively. }
\label{fig:Bound_Yuka}
\end{center}
\end{figure}

Under the constraints of Table~\ref{tab:nu_GF} and Table~\ref{tab:upper_exp}, we have sampled $10^9$ points with the ranges of involved parameters shown in Fig.~\ref{fig:Bound_Yuka}, where the black (red) points denote the NO (IO) neutrino mass.  Figs.~\ref{fig:Bound_Yuka}(a)--(c) show the allowed ranges of various products of Yukawa couplings in absolute values that will appear in the calculations of LFV processes, while Fig.~\ref{fig:Bound_Yuka}(d) shows those of $m_{N,S_1}$.  The plots indicate that the constrained Yukawa couplings in the model are not sensitive to the neutrino mass ordering.

%%%%%%%%%%%%%%%%%%%%%%%%%%%%%%%%%%%%%%%%%%%%%%%%%%%%%%%%%%%%
%%%%%%%%%%%%%%%%%%%%%%%%%%%%%%%%%%%%%%%%%%%%%%%%%%%%%%%%%%%%

%%%%%%%%%%%%%%%%%%%%%%%%%%%%%%%%%%%%%%%%%%%%%%%%%%%%%%%%%%%%
\subsection{Correlations among the LFV processes}
%%%%%%%%%%%%%%%%%%%%%%%%%%%%%%%%%%%%%%%%%%%%%%%%%%%%%%%%%%%%

In this subsection, we numerically discuss the influence of parameters on the rare LFV processes and the correlations among the considered LFV processes in detail.  Since the $\mu\to e\gamma$ decay is the most constraining LFV process at present and in foreseeable future, we first show its branching ratio dependence on the parameters and then investigate various correlations among the LFV processes.

\begin{figure}[phtb]
\begin{center}
\includegraphics[scale=0.6]{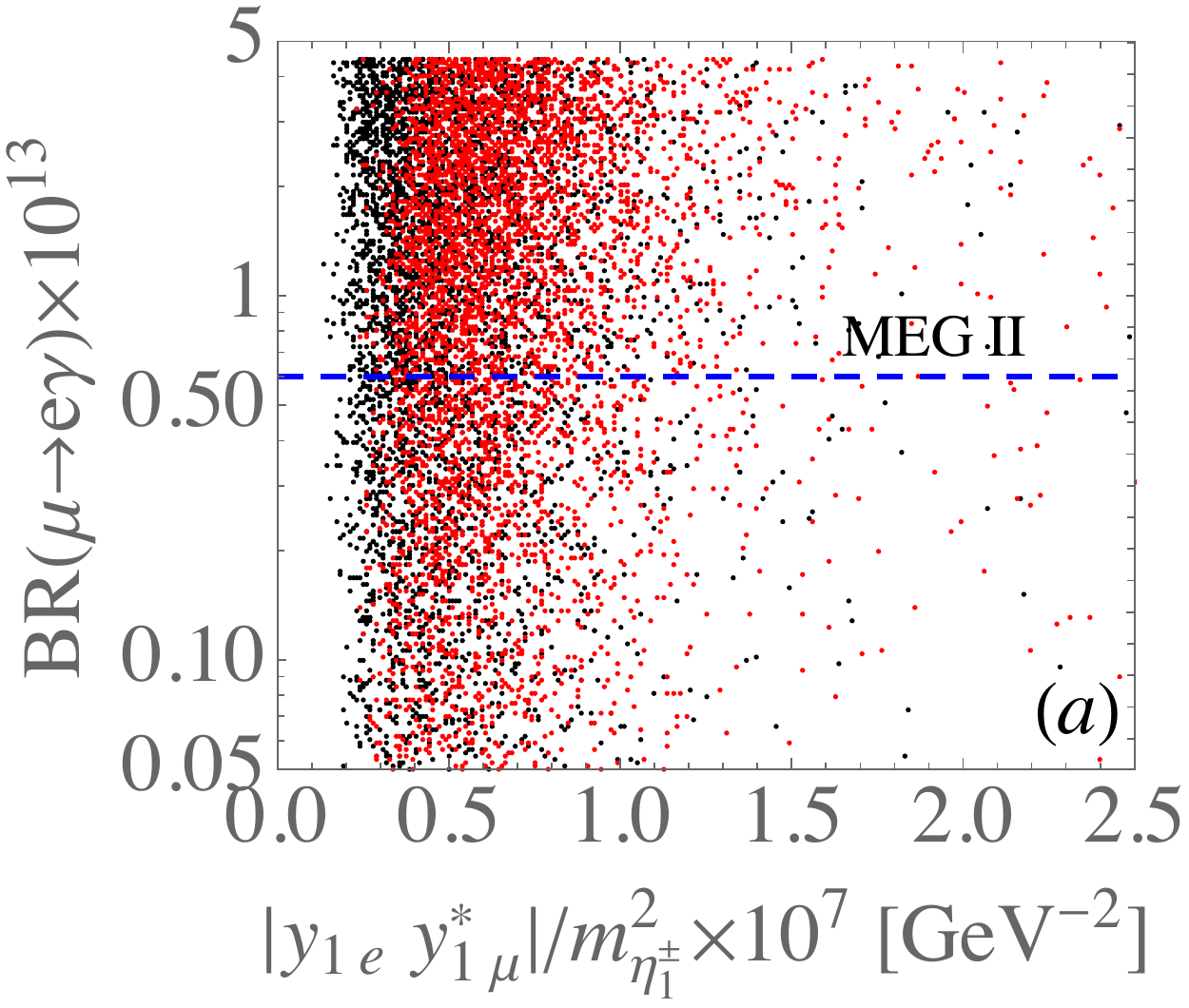}
\includegraphics[scale=0.6]{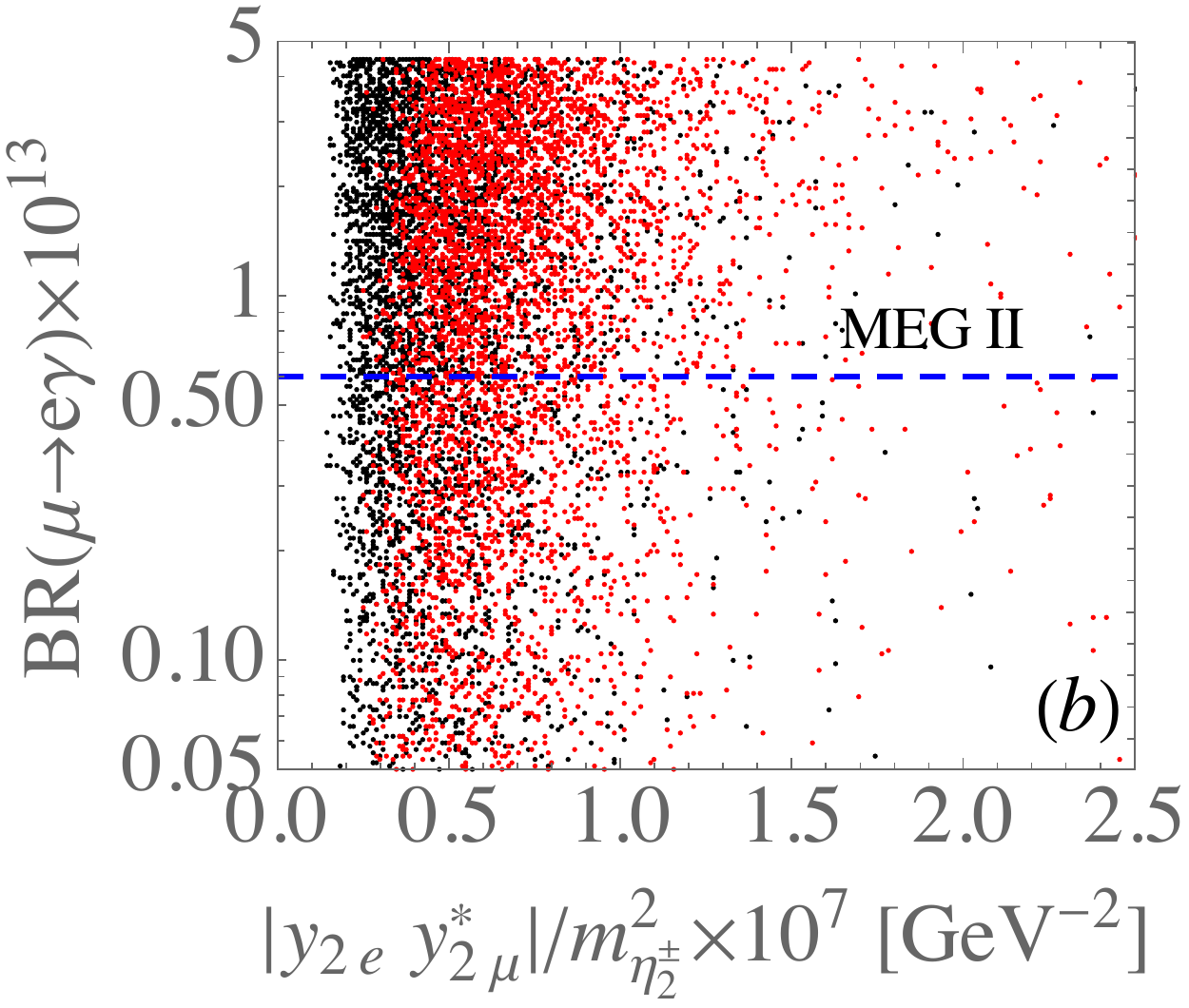}
\includegraphics[scale=0.6]{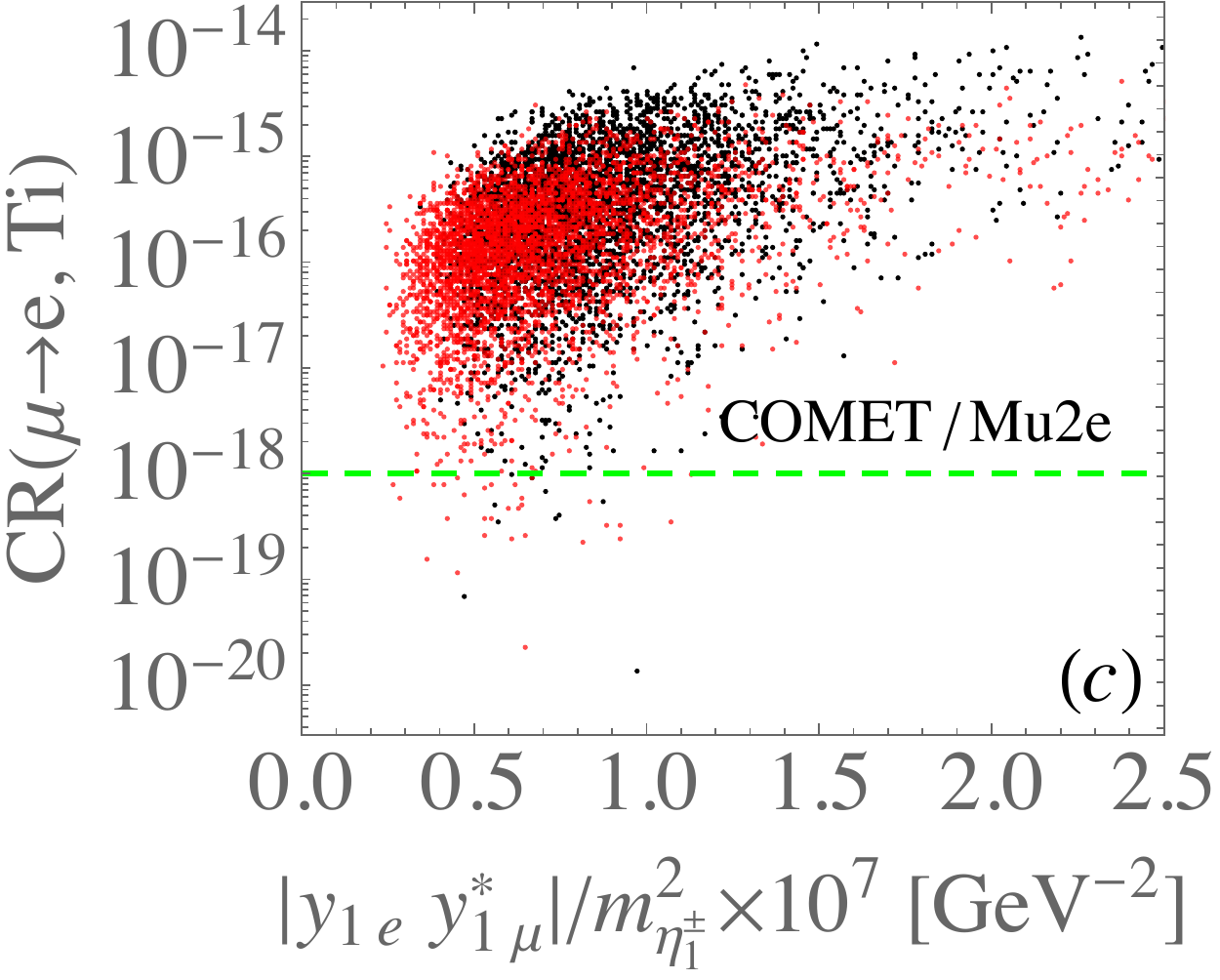}
\includegraphics[scale=0.6]{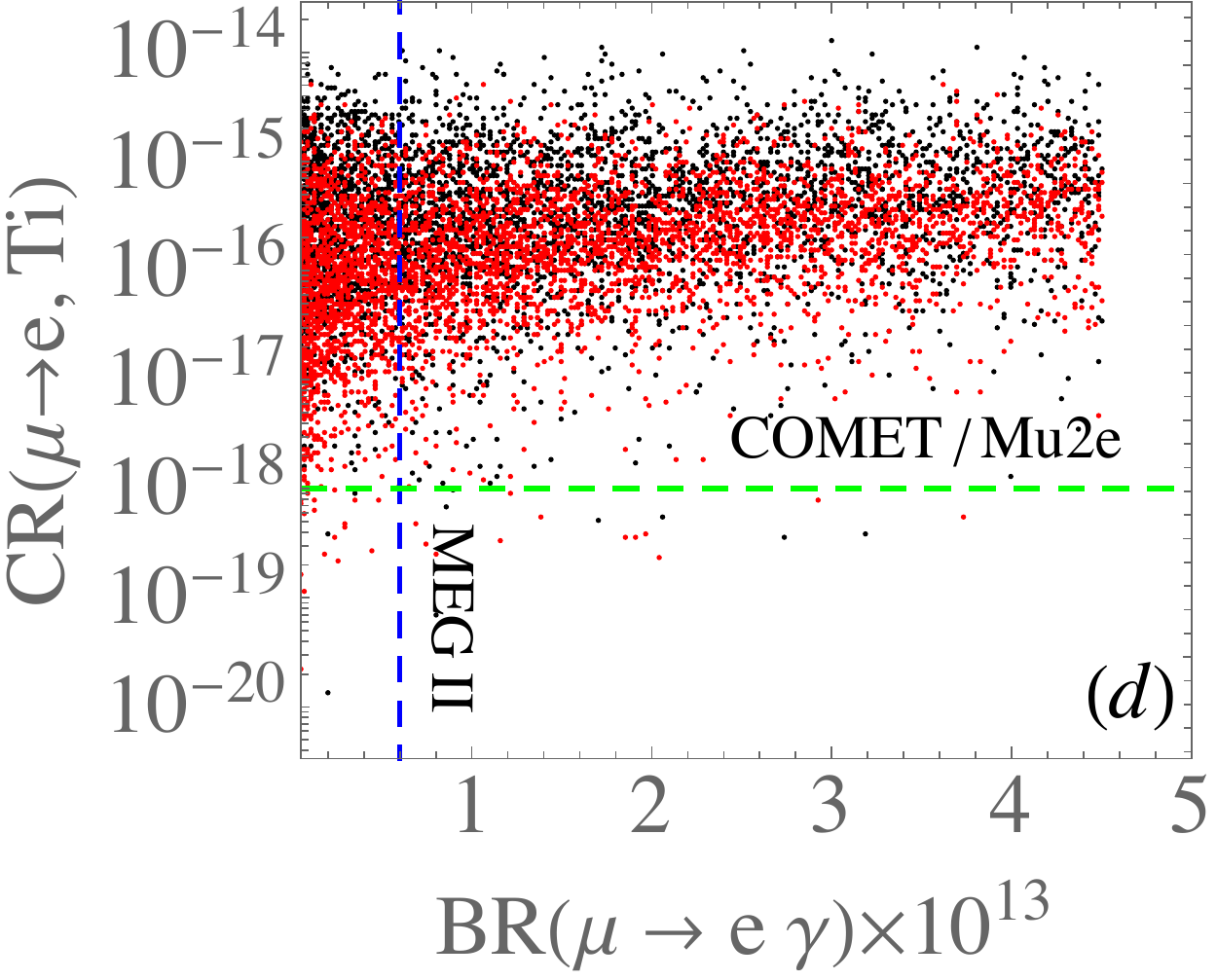}
\caption{ (a) and (b) $BR(\mu \to e \gamma)$ (in units of $10^{-13}$) as a function of $|y_{ie} y^*_{i \mu}|/m^2_{\eta^\pm_i}$; (c) $\mu-e$ conversion rate as a function of of $|y_{1e} y^*_{1 \mu}|/m^2_{\eta^\pm_1}$; and (d) correlation between $BR(\mu\to e \gamma)$ and $CR(\mu-e, {\rm Ti})$, where the dashed lines are the sensitivities of currently ongoing experiments.  The black and red points denote the results of NO and IO, respectively.}
\label{fig:LFV}
\end{center}
\end{figure}

According to Eq.~(\ref{eq:litoljga}), the branching ratio for $\mu\to e \gamma$ can be taken as a function of $|y_{ie} y^*_{i \mu}|/m^2_{\eta^\pm_i}$. Based upon the scanning results exhibited in Fig.~\ref{fig:Bound_Yuka},  we show the scatter plots for $BR(\mu \to e \gamma)$ as a function of $|y_{1e} y^*_{1 \mu}|/m^2_{\eta^\pm_i}$ and $|y_{2e} y^*_{2 \mu}|/m^2_{\eta^\pm_i}$ in Fig.~\ref{fig:LFV}(a) and (b), respectively, where the horizontal dashed line denotes the sensitivity of MEG II experiment~\cite{MEGII:2018kmf}.  Since the MEG II experiment can only probe $\mu\to e\gamma$ down to the level of $6\times 10^{-14}$, we have further restricted the model parameters to satisfy $BR(\mu\to e\gamma)\in (0.05,5)\times 10^{-13}$ in the plots.  As shown in the plots, the difference between the dependence on $|y_{1e} y^*_{1 \mu}|/m^2_{\eta^\pm_1}$ and that on $|y_{2e} y^*_{2 \mu}|/m^2_{\eta^\pm_2}$ is insignificant.  As given in Eqs.~(\ref{eq:CRmue}) and (\ref{eq:gqLV}), the $\mu-e$ conversion rate arises from the photon-penguin diagram in the model.  Thus, we show $CR(\mu-e,{\rm Ti})$ as a function of $|y_{1e} y^*_{1 \mu}|/m^2_{\eta^\pm_1}$ in Fig.~\ref{fig:LFV}(c), where the dashed line is the sensitivity of COMET~\cite{COMET:2018auw} and Mu2e~\cite{Diociaiuti:2020yvo} experiments.  As such, these experiments have the capability to probe most of the considered parameter space through the $\mu-e$ conversion rate.  The correlation between $BR(\mu\to e\gamma)$ and $CR(\mu\to e, {\rm Ti})$ is plotted in Fig.~\ref{fig:LFV}(d).  Again, when the measurement of $CR(\mu-e,{\rm Ti})\sim 10^{-18}$ achieves the expected sensitivity, the parameter space with $BR(\mu\to e \gamma) \gtrsim O(10^{-13})$ is mostly covered.  Hence, the $\mu-e$ conversion process has the potential to become the most stringent constraint among the LFV processes.

%%%%%%%%%%%%%%%%%%%%%%%%%%%%%%%%%%%%%%%%%%%%%%%%%%%%%%%%%%%%
%%%%%%%%%%%%%%%%%%%%%%%%%%%%%%%%%%%%%%%%%%%%%%%%%%%%%%%%%%%%

As stated earlier, in addition to the photon-penguin diagrams, the $\mu\to 3e$ decay can be induced by the box diagrams in the model. To exhibit the role of the box diagrams, the ratio of $BR(\mu\to 3e)$ purely from the photon-penguin contribution to $BR(\mu\to e \gamma)$ can be expressed as:
 \begin{equation}
 R_{3e/e\gamma}=\frac{BR(\mu\to 3e)_\gamma}{BR(\mu\to e\gamma)} \approx \frac{\alpha_{\rm em}}{8 \pi} \left[ \frac{|C^\gamma_{1e\mu}|^2}{|C^\gamma_{2e\mu}|^2} + \frac{16}{3} \ln \frac{m_\mu}{m_e} - \frac{22}{3} -4 \operatorname{Re}\left( \frac{C^\gamma_{1e\mu}}{C^\gamma_{2e\mu}}\right)\right]~. 
 \label{eq:r_muto3e_muega}
 \end{equation}
With $C^\gamma_{1e\mu}/C^\gamma_{2e\mu}\sim {\cal O}(1)$, the ratio in Eq.~(\ref{eq:r_muto3e_muega}) can be estimated to be $R_{3e/e\gamma} \sim 5 \times 10^{-3}$.  Using the constrained parameter values shown in Fig.~\ref{fig:Bound_Yuka}, indeed, we approximately obtain $R_{3e/e\gamma} \in (0.5, 5) \times 10^{-2}$.  We can conclude that if $BR(\mu\to 3e) > BR(\mu\to e\gamma)$ is observed in experiments, the enhancement should arise from other effects than from the photon-penguin diagrams.

\begin{figure}[pthb]
\begin{center}
\includegraphics[scale=0.45]{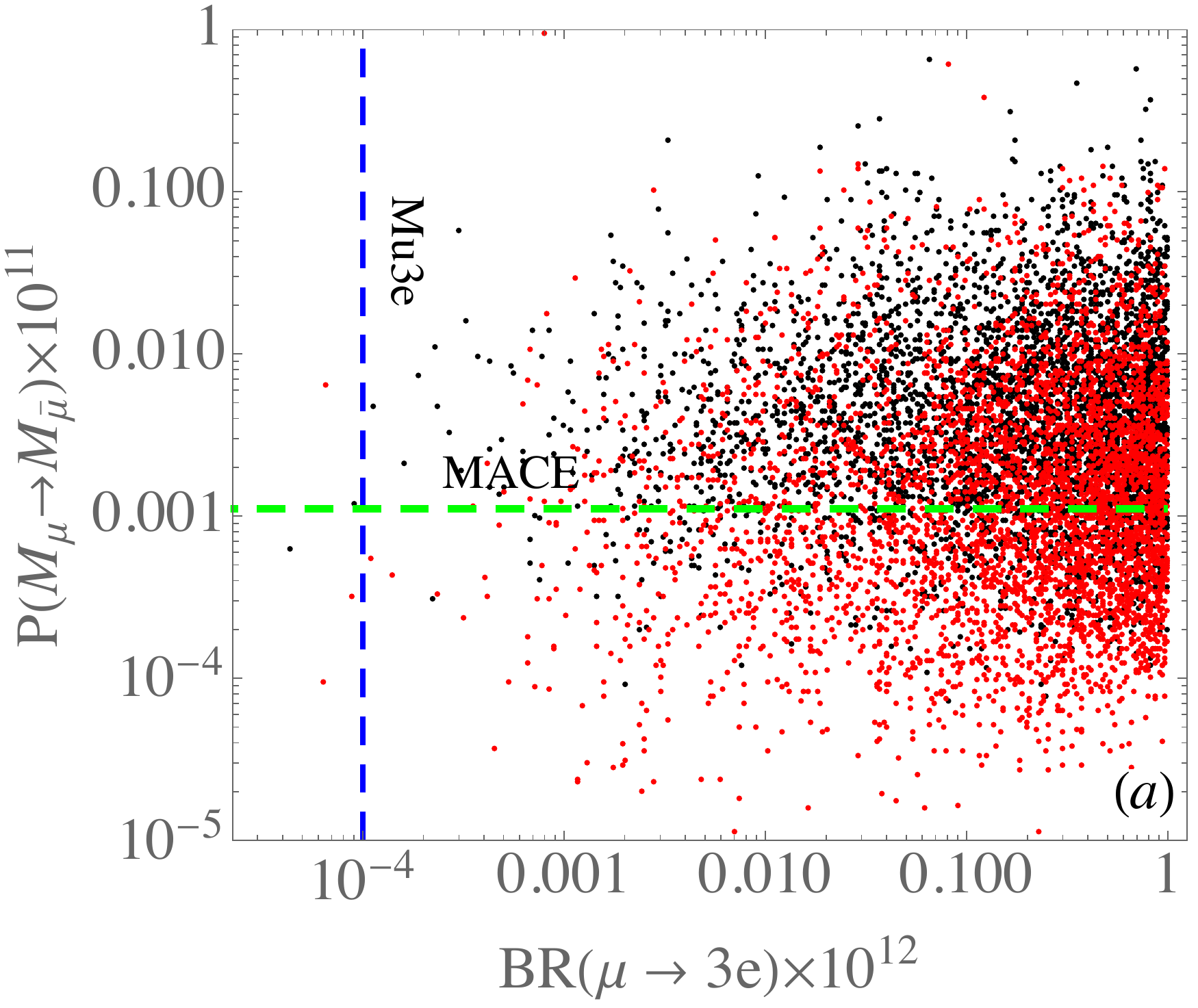}
\includegraphics[scale=0.45]{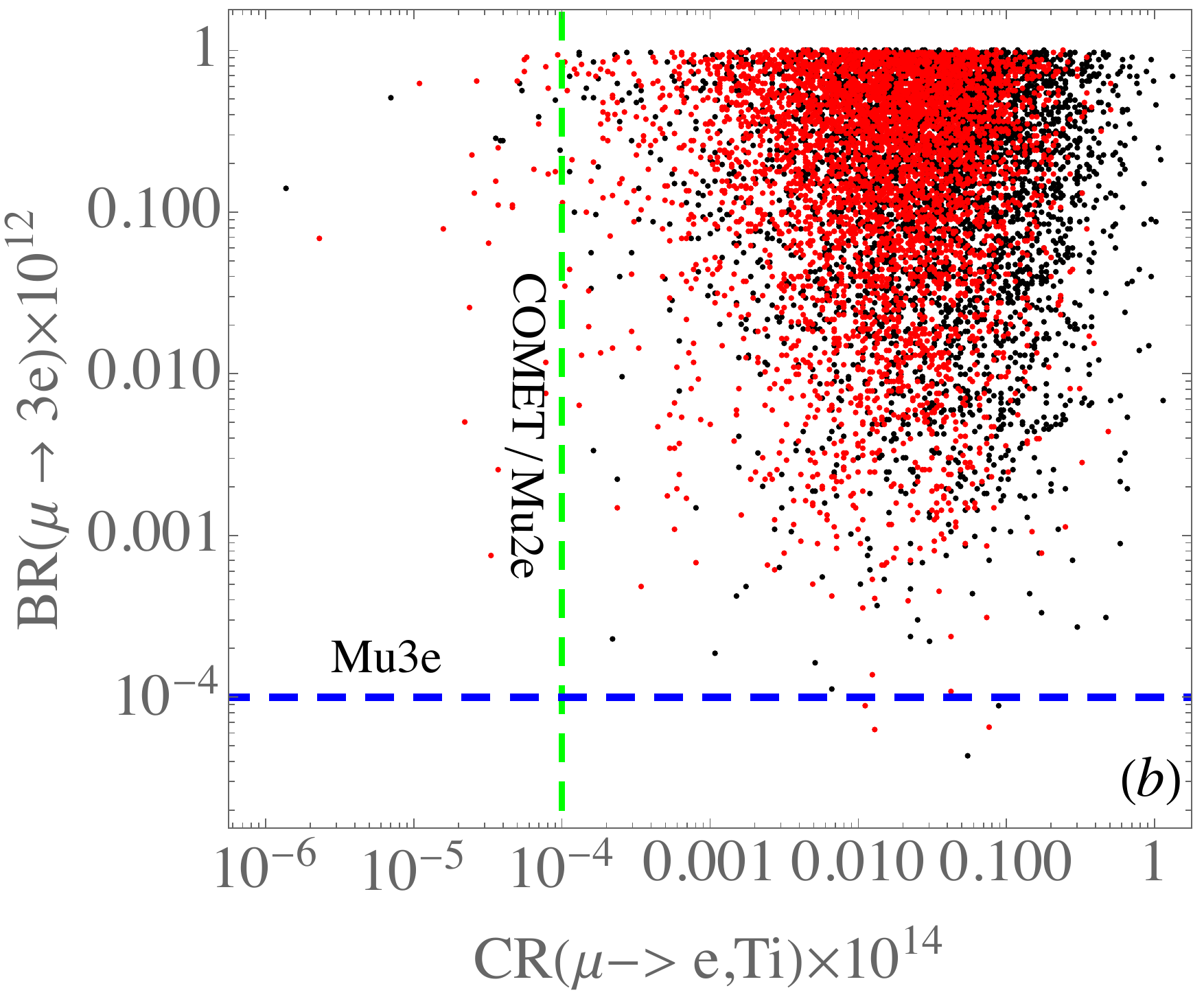}
\caption{Correlation between $BR(\mu\to 3e)$ and (a) $P(M_\mu-\overline M_\mu)$ and (b) $CR(\mu-e,{\rm Ti})$, where the dashed lines label the experimental sensitivities.  The black and red points denote the results of NO and IO, respectively.}
\label{fig:muonium-mu3e}
\end{center}
\end{figure}

If we assume that the photon-penguin contribution to $\mu\to 3 e$ is subleading, it is interesting to compare the $\mu\to 3e$ decay with the munoium oscillation, where both processes are dominated by the box diagrams.  Implementing the constrained parameter values shown in Fig.~\ref{fig:Bound_Yuka}  to the formulas in Eqs.~(\ref{eq:3ell}) and (\ref{eq:P_muonium}), we plot the correlation between $P(M_\mu-\overline M_\mu)$ and $BR(\mu\to 3e)$ in Fig.~\ref{fig:muonium-mu3e}(a), where the dashed lines label the sensitivities of Mu3e and MACE experiments.  It is seen that the predicted values of $P(M_\mu-\overline M_\mu)$ in the model are mostly below than the current experimental limit of $4.3 \times 10^{-12}$ when the $\mu\to 3 e$ decay is constrained to satisfy the current upper limit of $1.0 \times 10^{-12}$.  However, when $BR(\mu\to 3e)$ reaches the sensitivity of $10^{-16}$, the decay can cover most of the considered parameter space, {\it i.e.}, under the presumption that $BR(\mu\to e \gamma)\in (0.05, 5) 10^{-13}$.  We thus see that the two strictest constraints on the $\mu\to e$ transitions come from the $\mu-e$ conversion in nucleus and the $\mu\to 3 e$ decay.  Their correlation in the model can be found in Fig.~\ref{fig:muonium-mu3e}(b).  As such, if we do not see any evidence of the model in the LFV experiments, the highly sensitive measurements of $BR(\mu\to 3 e)$ and $\mu-e$ conversion rate severely constrain the $\mu\to e\gamma$ decay and munoium oscillation processes.

%%%%%%%%%%%%%%%%%%%%%%%%%%%%%%%%%%%%%%%%%%%%%%%%%%%%%%%%%%%%
%%%%%%%%%%%%%%%%%%%%%%%%%%%%%%%%%%%%%%%%%%%%%%%%%%%%%%%%%%%%

\begin{figure}[phtb]
\begin{center}
\includegraphics[scale=0.5]{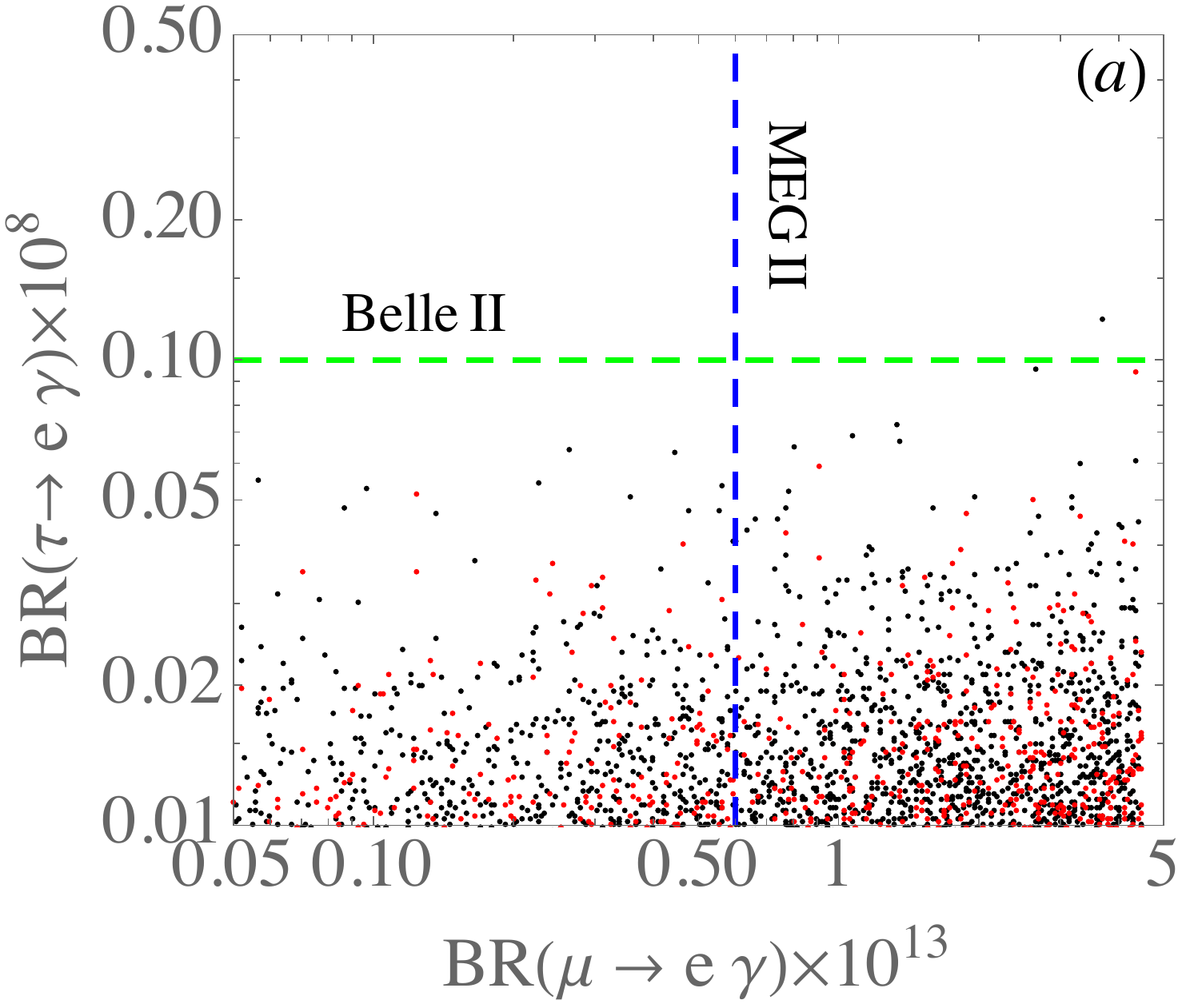}
\includegraphics[scale=0.5]{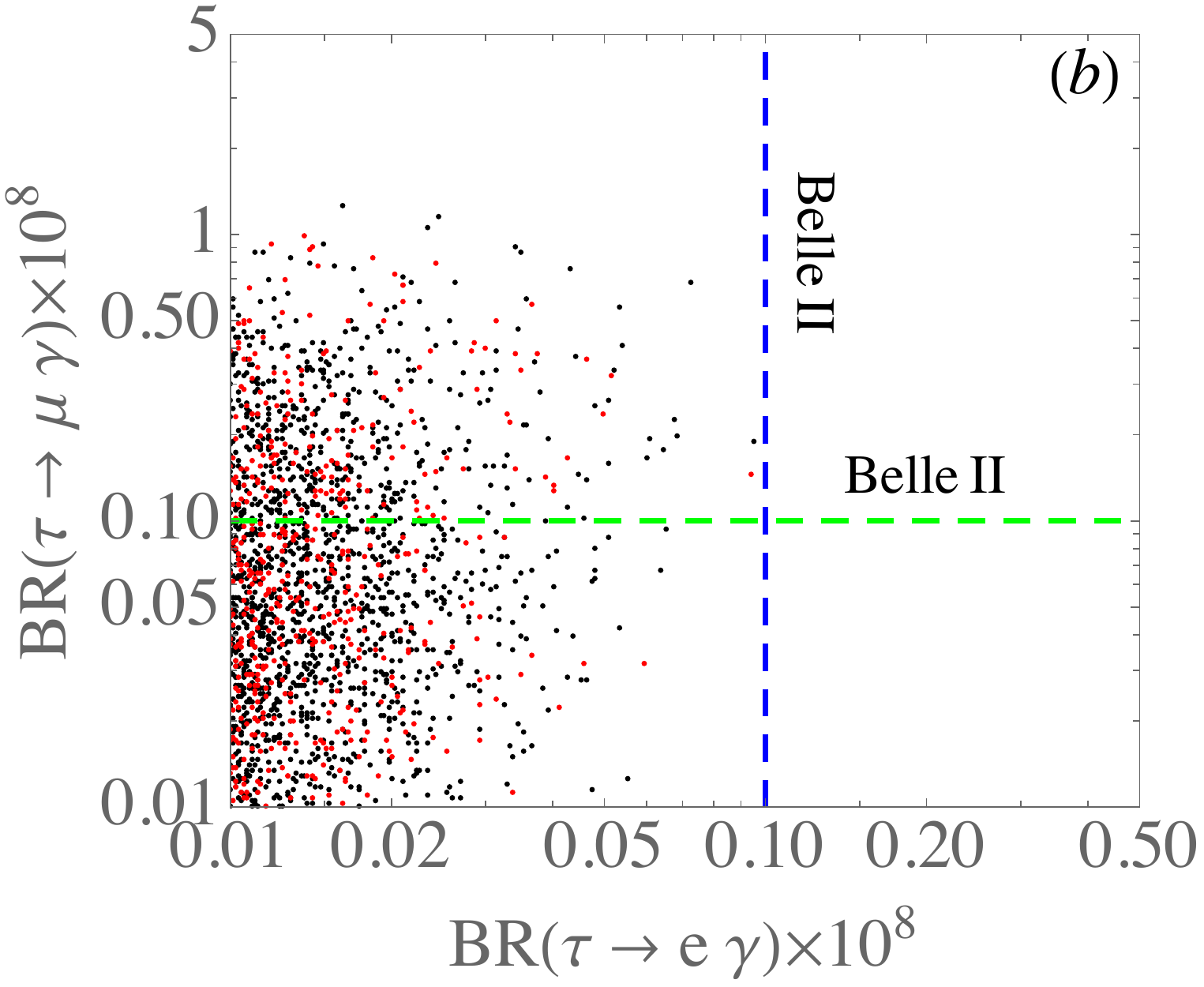}
\caption{ Correlation of BR in LFV processes: (a) $\tau\to e\gamma$ and $\mu\to e\gamma$, and (b) $\tau\to \mu\gamma$ and $\tau\to e\gamma$.  The black and red points denote the results of NO and IO, respectively. }
\label{fig:tautoellga}
\end{center}
\end{figure}

After discussing the rare $\mu\to e$ transition, we discuss in the following analysis the lepton flavor-changing effects on the heavy $\tau$-lepton decays, where the sensitivities assuming 50 ab$^{-1}$ of data at Belle II can reach~\cite{Belle-II:2018jsg}:
 \begin{align}
 \begin{split}
 BR(\tau \to \ell \gamma) & \sim 10^{-8} - 10^{-9}~, 
 \\
 BR(\tau\to  3 \ell) & \sim 10^{-9}-10^{-10}~,
 \end{split}
 \end{align}
with $\ell=e, \mu$.  Analogous to $\mu\to e \gamma$, the radiative $\tau$ decay also arises from the same types of diagrams.  Therefore, using Eq.~(\ref{eq:litoljga}) and the constrained parameter values, we plot the correlation between $BR(\tau\to e \gamma)$ and $BR(\mu\to e\gamma)$ in Fig.~\ref{fig:tautoellga}(a), where the dashed lines are the sensitivities of MEG II and Belle II experiments.  Since the resulting branching ratio for $\tau\to e\gamma$ is lower than the sensitivity of Belle II, it is difficult to observe the $\tau\to e\gamma$ decay in the model.  In Fig.~\ref{fig:tautoellga}(b), we show the correlation between $BR(\tau\to \mu \gamma)$ and $BR(\tau\to e\gamma)$.  It can be found that unlike the $\tau\to e \gamma$ decay, the branching ratio for $\tau\to \mu \gamma$ can be as large as $O(10^{-8})$.  As such, $\tau\to \mu \gamma$ serves as a good candidate to probe the new physics effects of the model in Belle II experiment.

%%%%%%%%%%%%%%%%%%%%%%%%%%%%%%%%%%%%%%%%%%%%%%%%%%%%%%%%%%%%
%%%%%%%%%%%%%%%%%%%%%%%%%%%%%%%%%%%%%%%%%%%%%%%%%%%%%%%%%%%%

\begin{figure}[phtb]
\begin{center}
\includegraphics[scale=0.5]{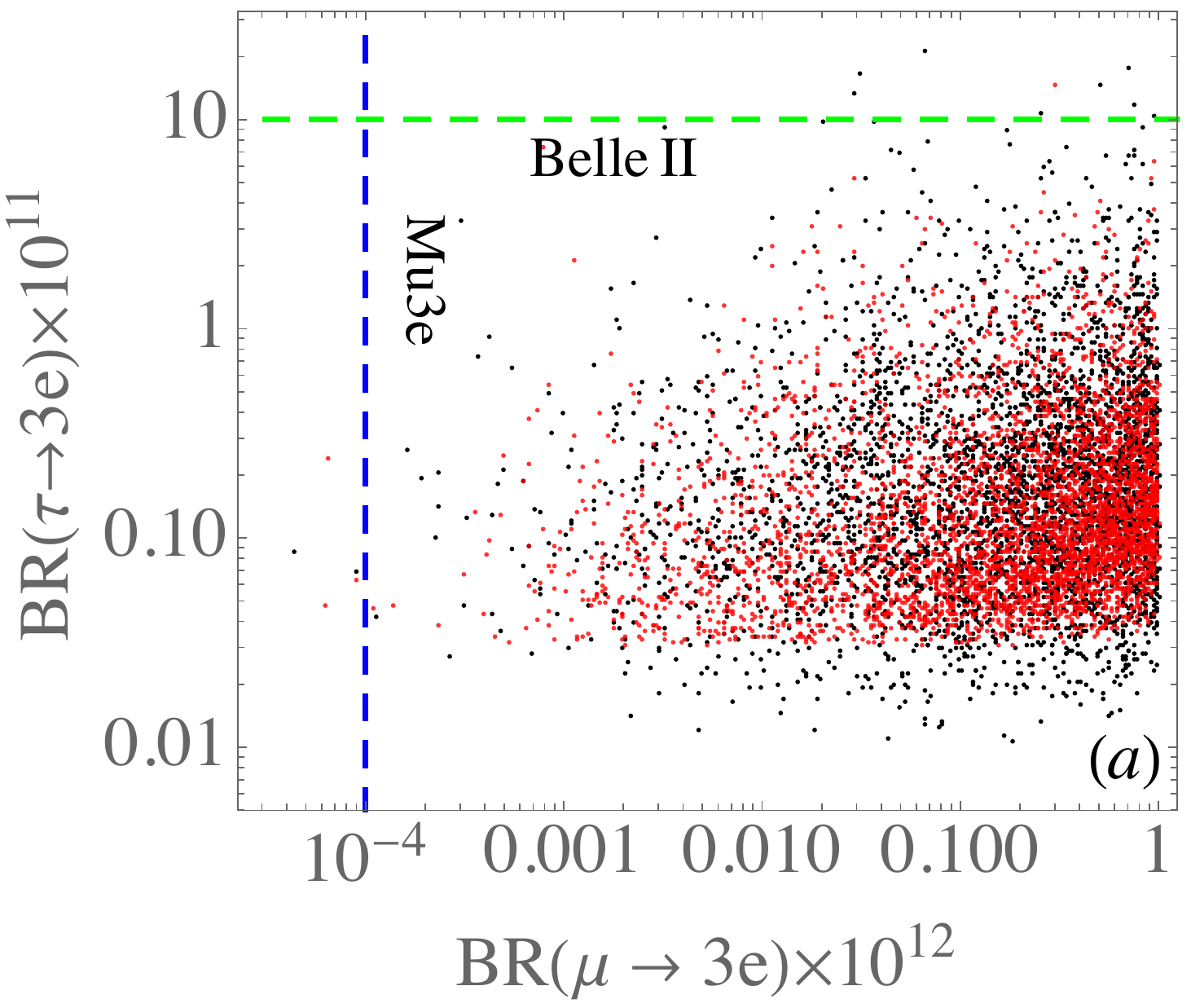}
\includegraphics[scale=0.5]{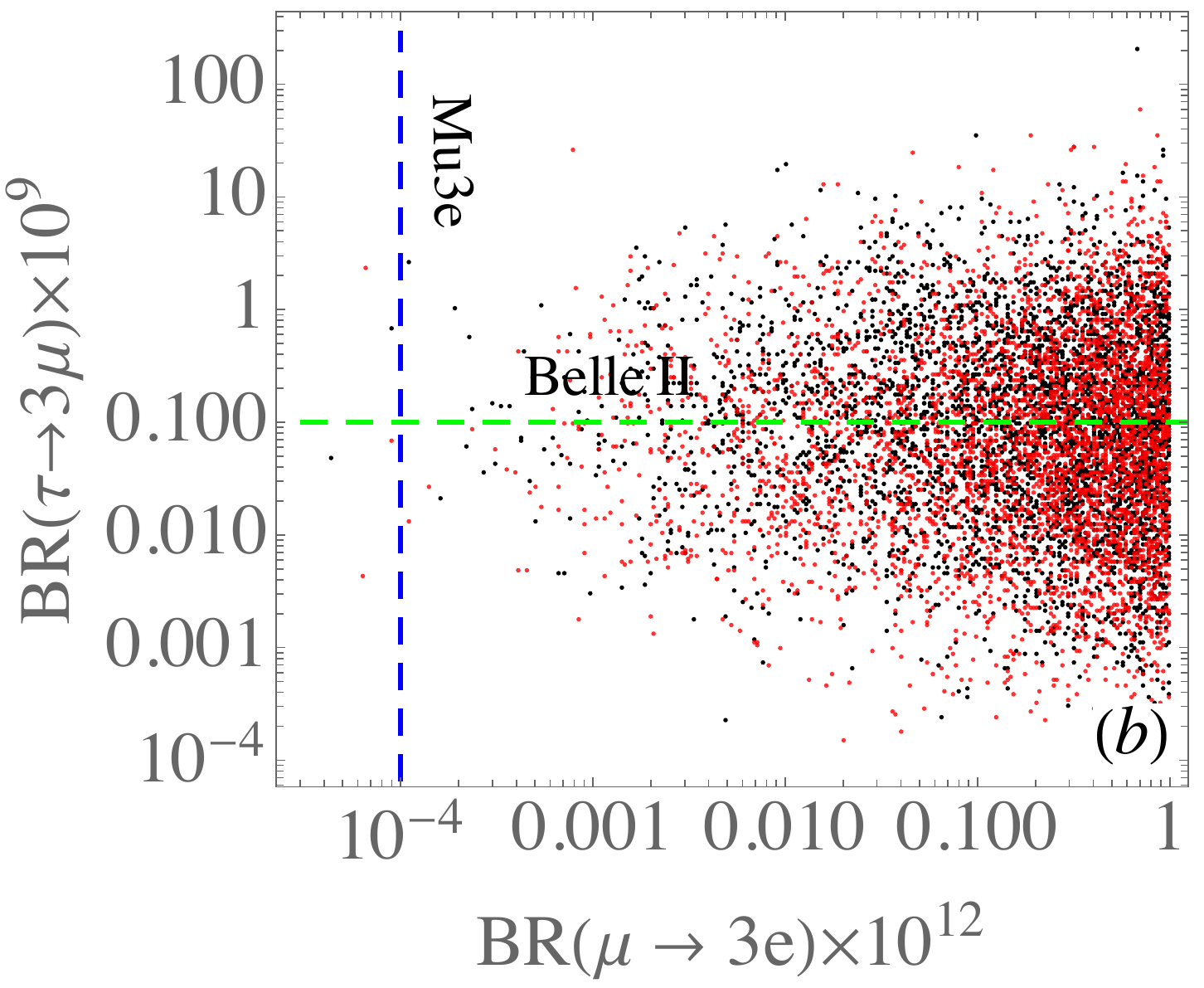}

\caption{ Correlation of BR between $\mu\to 3e$ and (a) $\tau\to 3e$ and (b) $\tau\to 3\mu$.  The black and red points denote the results of NO and IO, respectively.}
\label{fig:tauto3ell}
\end{center}
\end{figure}

As discussed earlier, although the $\ell_i \to 3\ell_j $ decay is induced by the photon-penguin and box diagrams, the effects of the latter are more dominant.  In order to reveal their contributions to $\tau\to 3e$ and $\tau\to 3\mu$, we show their branching ratio correlations with $\mu\to 3 e$ in Fig.~\ref{fig:tauto3ell}.  Similar to $\tau\to e \gamma$, it is difficult for Belle II to reach the predicted $BR(\tau\to 3e)$ in the model.  Nevertheless, the $\tau\to 3\mu$ decay is more promising to detect at Belle II because the value can reach $O(10^{-9}-10^{-10})$.

%%%%%%%%%%%%%%%%%%%%%%%%%%%%%%%%%%%%%%%%%%%%%%%%%%%%%%%%%%%%
%%%%%%%%%%%%%%%%%%%%%%%%%%%%%%%%%%%%%%%%%%%%%%%%%%%%%%%%%%%%

We finally make some remarks on the lepton $(g-2)$'s shown in Eq.~(\ref{eq:gm2}).  It is known that the lepton $(g-2)$ mediated by the charged scalar boson is usually negative.  Although the sign of $a_\mu$ in the model contradicts with that given by the recent E989 experiment at Fermilab~\cite{Abi:2021gix}, its value $|a_\mu|< 10^{-11}$ due to the $m^2_\mu$ dependence and $m_{\eta^\pm_i}\sim O(1)$~TeV.  Therefore, if the muon $g-2$ anomaly can be resolved within the SM~\cite{Borsanyi:2020mff}, the negative $a_\mu$ of the model will not cause a serious problem.  On the other hand, the electron $(g-2)$ in the model can be estimated as $|a_e|< 10^{-17}$ and is negligible.

%%%%%%%%%%%%%%%%%%%%%%%%%%%%%%%%%%%%%%%%%%%%%%%%%%%%%%%%%%%%
\subsection{DM relic density and DM-nucleon scattering cross section}
%%%%%%%%%%%%%%%%%%%%%%%%%%%%%%%%%%%%%%%%%%%%%%%%%%%%%%%%%%%%

In this subsection we discuss the DM phenomenology in our model.  Our DM candidate is the vector-like neutral fermion $N$ which interacts with the SM particles via the Yukawa couplings in Eq.~\eqref{eq:yukawa} and the $Z'$ exchange with the kinetic mixing effect.
In the estimate of relic density, the dominant DM annihilation processes are summarized as follows:
\begin{itemize}
\item $N \overline N \to Z' \to f_{SM} \overline f_{SM}, W^+W^-$.
\item $N \overline N \to Z' Z'$.
\item $\{ N  \overline N, NN, \overline N \overline N \} \to \{ \nu \overline \nu, \nu \nu, \overline \nu \overline \nu, \ell^+ \ell^- \}$, %through Yukawa interactions.
\end{itemize}
where $\ell^\pm$ and $f_{SM}$ denote the SM charged leptons and fermions, respectively.  The first two processes are mediated by the new gauge interactions, while the last channels rely mostly on the new Yukawa couplings.  We estimate the relic density of DM using {\tt micrOMEGAs 5.2.4}~\cite{Belanger:2014vza} implemented with the new interaction vertices in the model.

\begin{figure}[phtb]
\begin{center}
\includegraphics[scale=0.6]{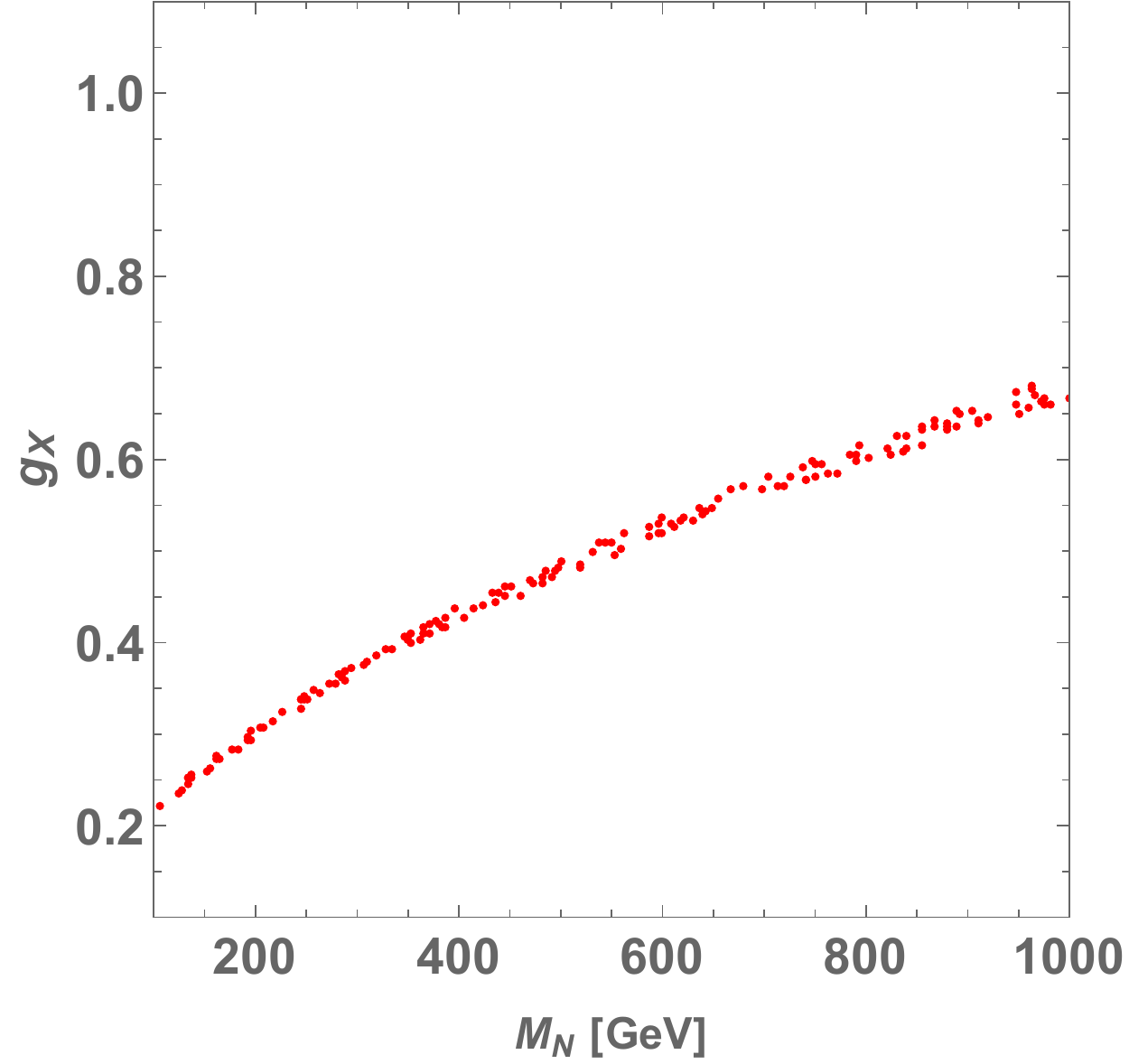} \ 
\includegraphics[scale=0.6]{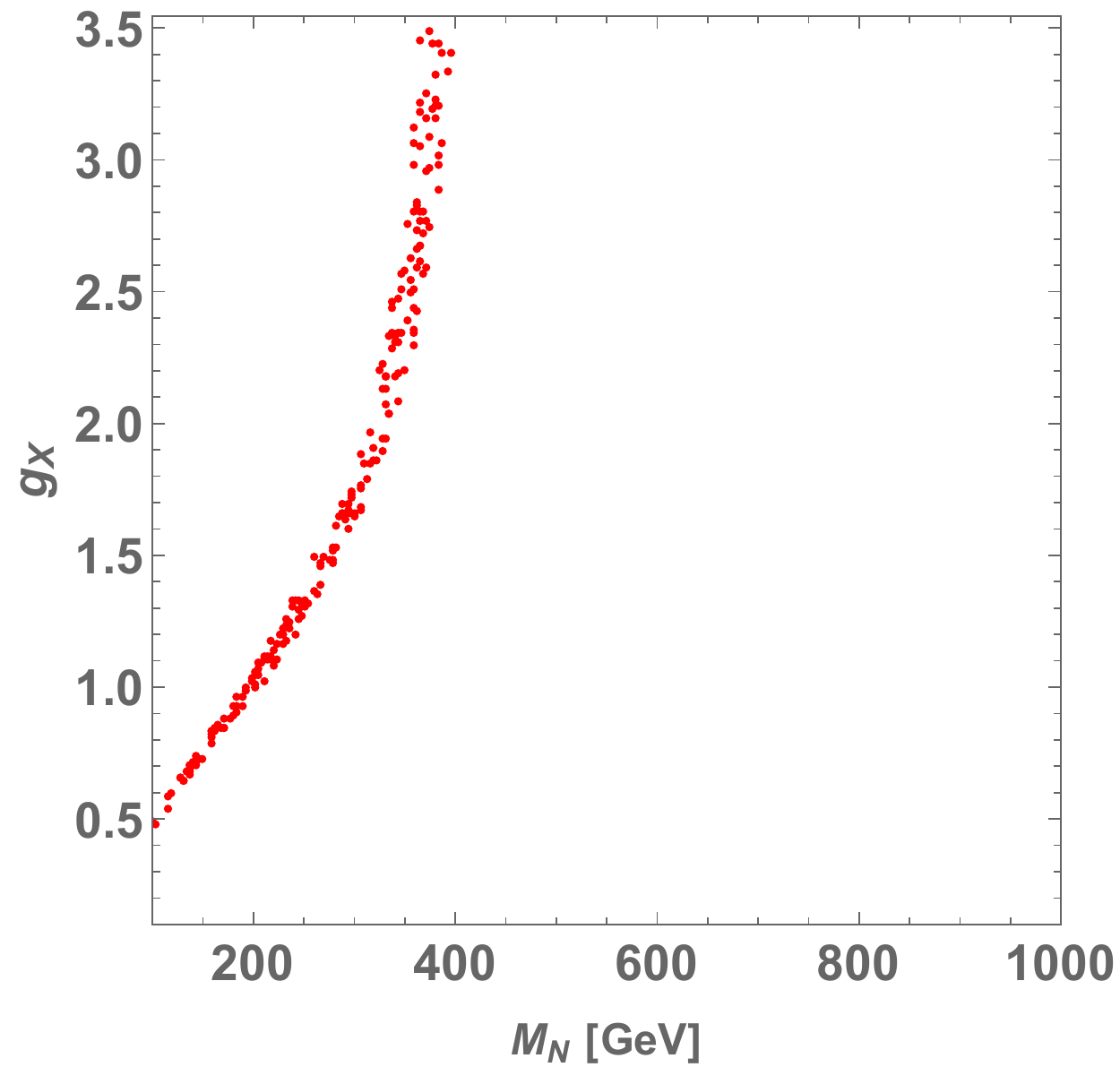}
\caption{Scatter plots of the parameter region in the $\{m_{N}$-$g_X\}$ plane that satisfies $0.11 < \Omega h^2 < 0.13$, considering only the $Z'$ interactions.  The Left and the right plots correspond to cases of $m_N = 2 m_{Z'}$ and $m_{Z'} = 2.025 m_N$, respectively. }
\label{fig:DM1}
\end{center}
\end{figure}

We first discuss the relic density by considering only $Z'$ interactions and choosing vanishing Yukawa couplings for the DM mass in range of 100~GeV to 1~TeV.  For illustration purposes, we consider two specific cases; (1) $m_N = 2 m_{Z'}$ where the $ N \overline N \to Z' Z'$ process is dominant, (2) $m_{Z'} = 2.025 m_N$ where the $N \overline N \to Z' \to \{f_{SM} \overline f_{SM}, W^+W^-\}$ processes are dominant.  The relic density is then estimated by scanning the values of $\{m_{N}, g_X\}$.  The left and right plots in Fig.~\ref{fig:DM1} show the parameter regions in the $m_{N}$-$g_X$ plane that give the relic density $0.11 < \Omega h^2 < 0.13$~\cite{PDG} for cases (1) and (2), respectively.  For case (1), the gauge coupling around $0.2 \lesssim g_X \lesssim 0.7$ can accommodate the observed relic density in the assumed DM mass range.  For case (2), on the other hand, a larger gauge coupling is required and the observed relic density can be explained only when $m_{N} \lesssim 400$~GeV imposing perturbative condition $g_X < \sqrt{4 \pi}$.  This behavior is due to that fact that the small kinetic mixing for the $s$-channel annihilation via $Z'$ exchange is suppressed.

%%%%%%%%%%%%%%%%%%%%%%%%%%%%%%%%%%%%%%%%%%%%%%%%%%%%%%%%%%%%
%%%%%%%%%%%%%%%%%%%%%%%%%%%%%%%%%%%%%%%%%%%%%%%%%%%%%%%%%%%%

\begin{figure}[phtb]
\begin{center}
\includegraphics[scale=0.7]{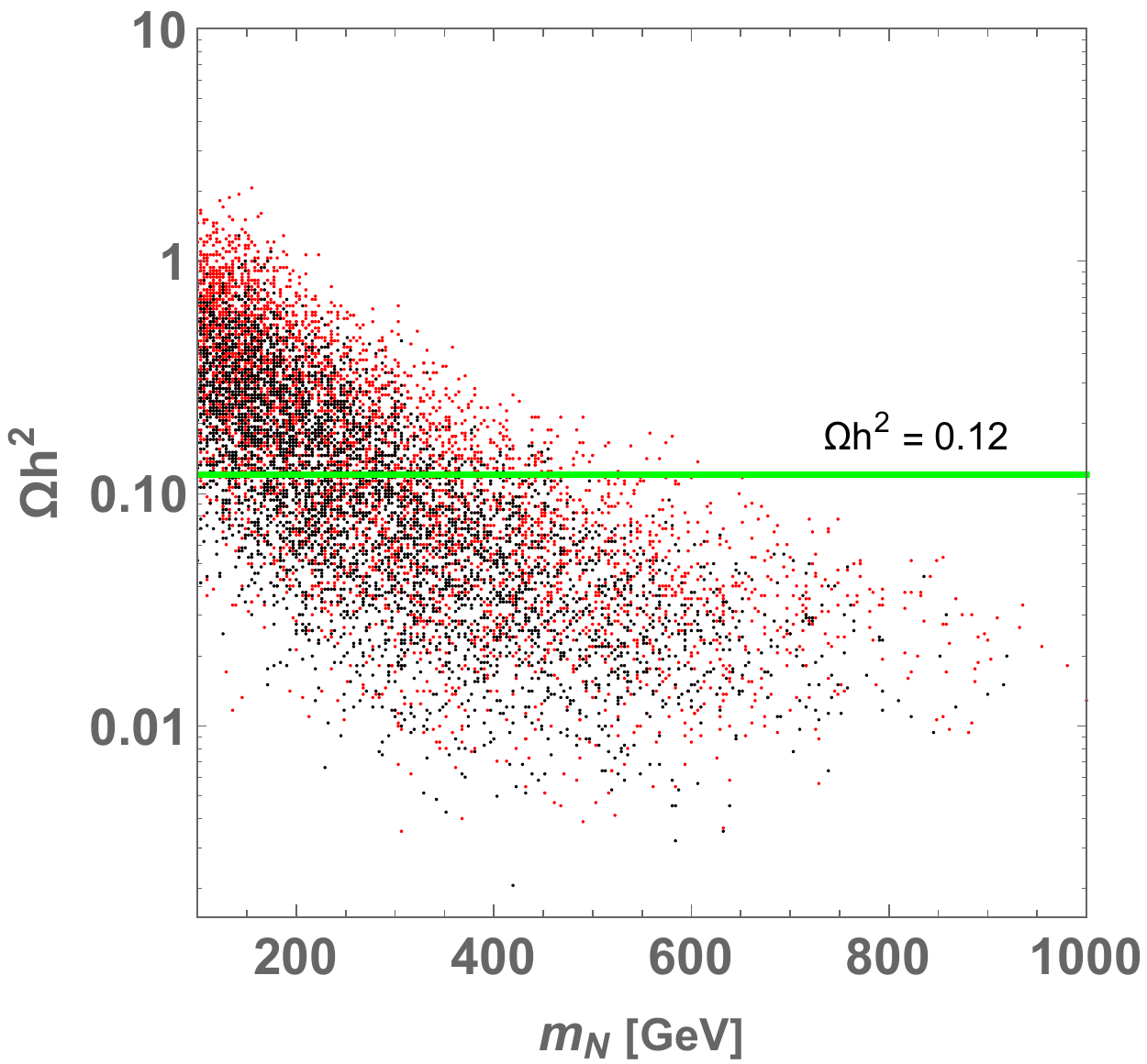}
\caption{Scatter plot of the relic density of $N$ for the allowed parameter sets with $g_X \to 0$.  The black and red points denote the results of NO and IO, respectively.}
\label{fig:DM2}
\end{center}
\end{figure}

Secondly, we discuss the relic density by use of the Yukawa couplings that satisfy the neutrino oscillation observations and flavor-changing constraints given in the previous subsections.  For illustration purposes, we consider a vanishing gauge coupling $g_X$ and focus on the effects of Yukawa couplings.  Fig.~\ref{fig:DM2} shows the DM relic density as a function of $m_N$ where the black and red points correspond to allowed parameter sets in the NO and IO scenarios, respectively.  We find that the observed relic density can be explained by the Yukawa interactions for $m_N \lesssim 650$~GeV and that it gets smaller than the observed value in the heavier DM region.  This is because the Yukawa couplings get larger in the heavier mass region, as required to fit the neutrino oscillation data.  We note in passing that it is possible to utilize the $Z'$ interactions to shift the scatter points downwards by turning on the gauge coupling $g_X$ and tuning its value and the $Z'$ mass.

%%%%%%%%%%%%%%%%%%%%%%%%%%%%%%%%%%%%%%%%%%%%%%%%%%%%%%%%%%%%
%%%%%%%%%%%%%%%%%%%%%%%%%%%%%%%%%%%%%%%%%%%%%%%%%%%%%%%%%%%%

\begin{figure}[phtb]
\begin{center}
\includegraphics[scale=0.7]{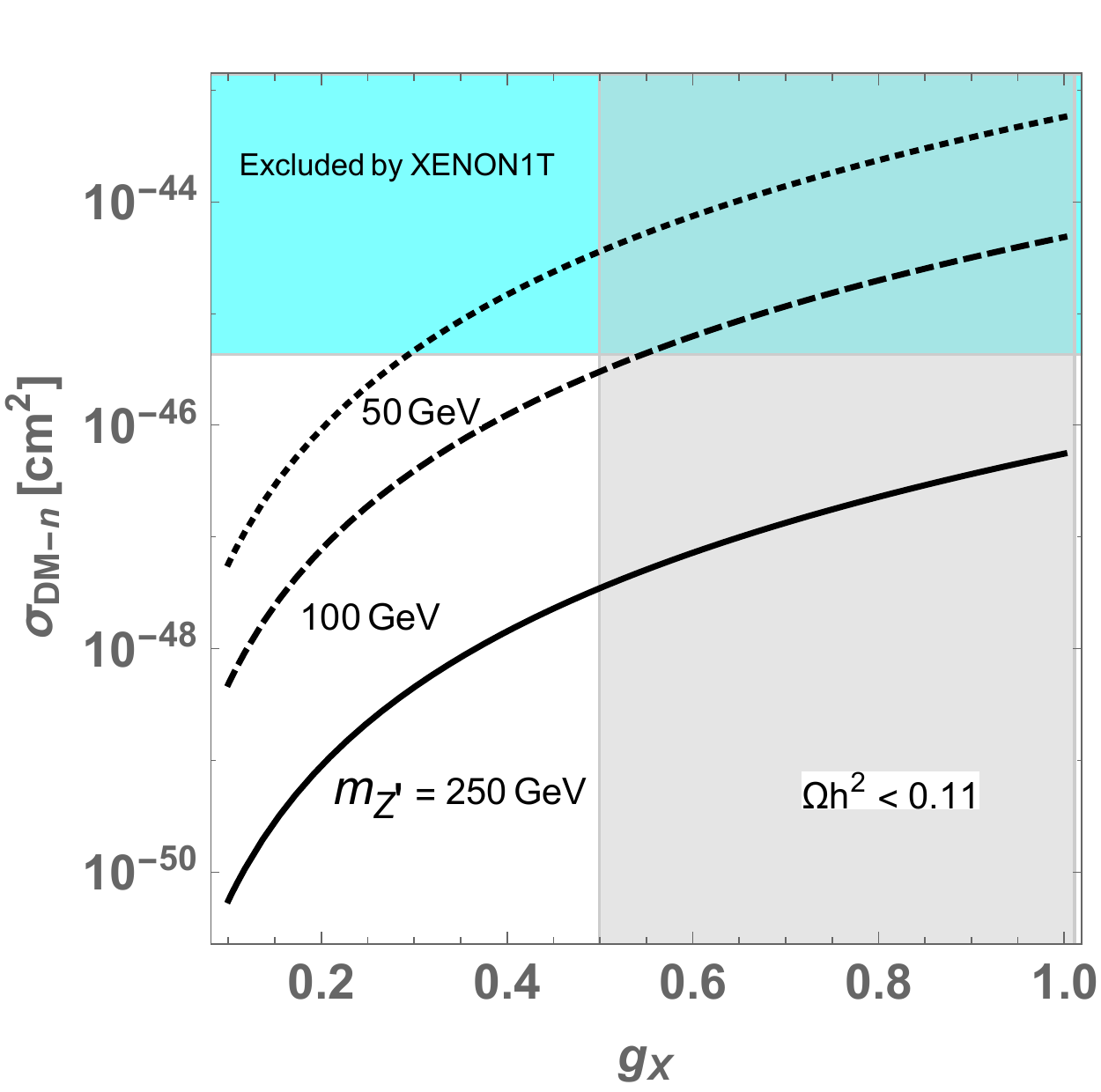}
\caption{DM-neutron scattering cross section as a function of $g_X$ for $m_{Z'} = \{50, 100, 250\}$~GeV where ray region indicates $g_X$ region providing $\Omega h^2 \lesssim 0.11$ and 
light blue region indicate excluded region by {\it XENON1T}, for $m_N = 500$ GeV.}
\label{fig:DD}
\end{center}
\end{figure}

Finally we discuss the DM-nucleon scattering cross section via the $Z'$ exchange.  Here we estimate the cross section in the non-relativistic limit and obtain 
\begin{align}
\begin{split}
\sigma_{n(p)} \simeq &~ \frac{g_X^2}{2 \pi} \frac{\mu_{n(p)}^2}{m^4_{Z'}} \left[ \left(C_{L}^{n(p)} \right)^2 + \left(C_{R}^{n(p)} \right)^2 + C_L^{n(p)} C_{R}^{n(p)} \right]
~, 
\\
\mbox{with } & C_{L(R)}^{n} = C_{L(R)}^u + 2 C_{L(R)}^d ~, \quad C_{L(R)}^{p} = 2C_{L(R)}^u +  C_{L(R)}^d
~,
\end{split}
\end{align}
where $n~(p)$ stands for the neutron~(proton), $\mu_{n(p)} \equiv m_N m_{n(p)}/(m_N + m_{n(p)})$ and $C_{L(R)}^f$ is given in Eq.~\eqref{eq:Zpff}. 
We show in Fig.~\ref{fig:DD} the DM-neutron scattering cross section as a function of $g_X$ for $m_{Z'} = \{50, 100, 250\}$~GeV and $m_N = 500$~GeV.  The gray region has $\Omega h^2 \lesssim 0.11$ and the cyan region is excluded by XENON1T~\cite{XENON:2018voc}.  In making this plot, we assume the dominance $Z'$ interactions and ignore the Yukawa couplings.  The DM-proton scattering cross section is very similar and thus omitted here.

%%%%%%%%%%%%%%%%%%%%%%%%%%%%%%%%%%%%%%%%%%%%%%%%%%%%%%%%%%%%
%%%%%%%%%%%%%%%%%%%%%%%%%%%%%%%%%%%%%%%%%%%%%%%%%%%%%%%%%%%%

%%%%%%%%%%%%%%%%%%%%%%%%%%%%%%%%%%%%%%%%%%%%%%%%%%%%%%%%%%%%
\section{Summary} \label{sec:summary}
%%%%%%%%%%%%%%%%%%%%%%%%%%%%%%%%%%%%%%%%%%%%%%%%%%%%%%%%%%%%

Based upon the concept of radiative neutrino mass in scotogenic model proposed in~\cite{Ma:2006km}, we have studied the possibility for   the Majorana neutrino mass to arise from an unbroken Stueckelberg $U(1)_X$ gauge model.  It is found that although we need two inert Higgs doublets to generate the neutrino mass, just one extra vector-like singlet fermion is sufficient to fit the neutrino data while in contrast at least two right-handed singlet fermions are required in the Ma model.  Using the proposed model, we have studied its implications on the low-energy lepton flavor violating (LFV) processes.

We have found that in the model, the resulting $BR(\mu\to e\gamma)$ can fit the currently experimental upper limit.  However, the planned sensitivities of the experiments on the $\mu-e$ conversion and the $\mu\to 3e$ decay in COMET/Mu2e and Mu3e can cover most of the parameter space that can lead to $BR(\mu\to e\gamma)\sim 6 \times 10^{-14}$, the designed sensitivity of MEG II experiment.  Hence, the $\mu-e$ conversion in nucleus and $\mu\to 3 e$ will be the most promising processes to detect the new physics in the $\mu\to e$ transitions.

In heavy lepton decays, although the predicted $BR(\tau\to e \gamma)$ and $BR(\tau\to 3e)$ are lower than $O(10^{-9})$ and $O(10^{-10})$, the branching ratios for $\tau\to \mu \gamma$ and $\tau\to 3\mu$ in the model can cover the range from $O(10^{-8})$ to $O(10^{-9})$ and $O(10^{-10})$, respectively, and can be probed by Belle II experiment.

We have discussed the dark matter (DM) phenomenology, including estimates of the relic density and the DM-nucleon scattering cross section for the DM mass in the range of 100~GeV to 1~TeV.  The relic density has been estimated for two cases: (i) the $Z'$ interaction is dominant (ii) the Yukawa interaction is dominant.  In case (i), the observed relic density can be explained with $0.2 \lesssim g_X \lesssim 0.7$ when DM annihilates into $Z'Z'$ while a larger gauge coupling is required when DM annihilate into the SM particles via the $s$-channel $Z'$ exchange.  In case (ii), we make use of the parameter sets obtained from neutrino oscillation observations and flavor-changing constraints.  We find that the relic density can be explained for a DM mass less than around 650~GeV.  The DM-nucleon cross section is obtained by considering the $Z'$ exchange process and evaluated by fixing some parameters.  We have found that it is possible to avoid constraints from direct detection while satisfying the relic density by properly choosing the gauge coupling and $Z'$ mass.

\section*{Acknowledgments}

This work was supported in part by the Ministry of Science and Technology, Taiwan under the Grant Nos.~MOST-110-2112-M-006-010-MY2 and MOST-108-2112-M-002-005-MY3.  
The work is also supported by the Fundamental Research Funds for the Central Universities (T.~N.).

%%%%%%%%%%%%%%%%%%%%%%%%%%%%%%%%%%%%%%%%%%%%%%%%%%%%%%%%%%%%

\end{document}